\documentclass[preprint,showpacs,preprintnumbers,amsmath,amssymb,
superscriptaddress,groupedaddress]{revtex4}

\usepackage{graphicx}
\usepackage{dcolumn}
\usepackage{bm}

\begin{document}

\title{Delocalization of wave packets in disordered nonlinear chains}

\author{Ch.~Skokos} \affiliation{Max Planck Institute for
the Physics of Complex Systems, N\"othnitzer Str. 38, D-01187 Dresden,
Germany}
\author{D.~O.~Krimer} \affiliation{Max Planck
Institute for the Physics of Complex Systems, N\"othnitzer Str. 38, D-01187
Dresden, Germany}
\author{S.~Komineas}
\affiliation{Department of Applied Mathematics, University of Crete, GR-71409
Heraklion, Crete, Greece}
\author{S.~Flach} \affiliation{Max Planck Institute for the Physics of Complex
Systems, N\"othnitzer Str. 38, D-01187 Dresden, Germany}

\date{\today}

\begin{abstract}
We consider the spatiotemporal evolution of a wave packet in disordered
nonlinear Schr\"odinger and anharmonic oscillator chains.  In the absence of
nonlinearity all eigenstates are spatially localized with an upper bound on
the localization length (Anderson localization).  Nonlinear terms in the
equations of motion destroy Anderson localization due to nonintegrability and
deterministic chaos.  At least a finite part of an initially localized
wave packet will subdiffusively spread without limits. We analyze the details
of this spreading process.  We compare the evolution of single site, single
mode and general finite size excitations, and study the statistics of
detrapping times.  We investigate the properties of mode-mode resonances,
which are responsible for the incoherent delocalization process.
\end{abstract}

\maketitle

\section{Introduction}
The normal modes (NMs) of a $d=1$--dimensional linear system with uncorrelated
random potential are spatially localized (Anderson localization).  Therefore
any wave packet, which is initially localized, remains localized for all time
\cite{PWA58}. Note that NMs correspond to single particle eigenstates of
related quantum systems.

When nonlinearities are added, NMs interact with each other \cite{GrKiv92}.
Recently, experiments were performed on light propagation in spatially random
nonlinear optical media \cite{Exp, Exp2} and on Bose-Einstein condensate
expansions in random optical potentials \cite{BECEXP}, which serve as
realizations of such cases.

Numerical studies of wave packet propagation in several models showed that the
second moment of the norm/energy distribution grows subdiffusively in time
as $t^{\alpha}$ \cite{Shep93,Mol98,PS08,Shep08}, with $\alpha \approx 1/3$ for
$d=1$. Reports on partial localization were published as well \cite{kkfa08}.

In a recent letter the mechanisms of spreading and localization were studied
for $d=1$, with initial excitations being localized on a single site
\cite{fks08}.  A theoretical explanation of the exponent $\alpha=1/3$ was
obtained, consistently assuming that the internal dynamics of a wave packet is
chaotic, leading to a partial dephasing of the NMs. The argumentation was
based on the possibility of a pair of wave packet modes being able to
resonantly interact with each other.  Among other results, the case of weak
nonlinearity showed that wave packets localize according to the linear dynamics
on long but finite time scales, with subsequent detrapping.  In the present
work, we extend this study to single mode excitations, and more general
 excitations of width $L$.  We study the details of the detrapping
process, and measure the statistical properties of detrapping times. We study
the particularities of resonant interaction between modes, mediated by the
nonlinearity. We give details on the used integration schemes, and perform
extensive tests which demonstrate that the observed effects are not affected
by roundoff errors. We argue that the spreading is inherently induced by the
nonintegrability of the system.

\section{Models}
\label{sec.model}

We study two models of one--dimensional lattices.

\subsection{Nonlinear Schr\"odinger lattice}

The Hamiltonian of the disordered discrete nonlinear Schr\"odinger equation
(DNLS)
\begin{equation}
\mathcal{H}_{D}= \sum_{l} \epsilon_{l}
|\psi_{l}|^2+\frac{\beta}{2} |\psi_{l}|^{4}
- (\psi_{l+1}\psi_l^*  +\psi_{l+1}^* \psi_l)
\label{RDNLS}
\end{equation}
with complex variables $\psi_{l}$, lattice site indices $l$ and nonlinearity
strength $\beta \geq 0$.   The random on-site energies $\epsilon_{l}$
are chosen uniformly from the interval
$\left[-\frac{W}{2},\frac{W}{2}\right]$, with $W$ denoting the disorder
strength.  The equations of motion are generated by $\dot{\psi}_{l} = \partial
\mathcal{H}_{D}/ \partial (i \psi^{\star}_{l})$:
\begin{equation}
i\dot{\psi_{l}}= \epsilon_{l} \psi_{l}
+\beta |\psi_{l}|^{2}\psi_{l}
-\psi_{l+1} - \psi_{l-1}\;.
\label{RDNLS-EOM}
\end{equation}
Eqs.~(\ref{RDNLS-EOM}) conserve the energy (\ref{RDNLS}) and the norm $S
= \sum_{l}|\psi_l|^2$.  We note that varying the norm of an
initial wave packet is strictly equivalent to varying $\beta$, therefore we
choose $S=1$.  Eqs.~(\ref{RDNLS}) and (\ref{RDNLS-EOM}) are derived e.~g.~when
describing two-body interactions in ultracold atomic gases on an optical
lattice within a mean field approximation \cite{oberthaler}, but also when
describing the propagation of light through networks of coupled optical
waveguides in Kerr media \cite{KivsharAgrawal}.

For $\beta=0$ Eq.~(\ref{RDNLS}) with $\psi_{l} = A_{l}
\exp(-i\lambda t)$
is reduced to the linear eigenvalue problem
\begin{equation}
\lambda A_{l} = \epsilon_{l} A_{l}
- A_{l-1}-A_{l+1}\;.
\label{EVequation}
\end{equation}
The normalized eigenvectors $A_{\nu,l}$ ($\sum_l A_{\nu,l}^2=1)$ are the NMs,
and the eigenvalues $\lambda_{\nu}$ are the frequencies of the NMs.  The width
of the eigenfrequency spectrum $\lambda_{\nu}$ of (\ref{EVequation}) is
$\Delta_D=W+4$ with $\lambda_{\nu} \in \left[ -2 -\frac{W}{2}, 2 + \frac{W}{2}
\right] $.

The asymptotic spatial decay of an eigenvector is given by $A_{\nu,l} \sim
{\rm e}^{-l/\xi(\lambda_{\nu})}$ where $\xi(\lambda_{\nu}) \leq \xi(0) \approx
100/W^2$ is the localization length \cite{KRAMER}. The NM participation number
$p_{\nu} = 1/\sum_l A_{\nu,l}^4$ characterizes the spatial extend
(localization volume) of the NM. It is distributed around the mean value
$\overline{p_{\nu}}\approx 3.6 \xi(\lambda_{\nu})$ with variance $\approx 1.3
\xi(\lambda_{\nu})$ \cite{MIRLIN}.  The average spacing of eigenvalues of NMs
within the range of a localization volume is therefore $\overline{\Delta
\lambda}_D \approx \Delta_D / \overline{p_{\nu}} \approx \Delta_D W^2 /360 $.
The two scales $ \overline{\Delta \lambda}_D \leq \Delta_D $ determine the
packet evolution details in the presence of nonlinearity.

The equations of motion of (\ref{RDNLS})  in normal mode space read
\begin{equation}
i \dot{\phi}_{\nu} = \lambda_{\nu} \phi_{\nu} + \beta \sum_{\nu_1,\nu_2,\nu_3}
I_{\nu,\nu_1,\nu_2,\nu_3} \phi^*_{\nu_1} \phi_{\nu_2} \phi_{\nu_3}\;
\label{NMeq}
\end{equation}
with the overlap integral
\begin{equation}
I_{\nu,\nu_1,\nu_2,\nu_3} =
\sum_{l} A_{\nu,l} A_{\nu_1,l}
A_{\nu_2,l} A_{\nu_3,l}\;.
\label{OVERLAP}
\end{equation}
The variables $\phi_{\nu}$ determine the complex time-dependent amplitudes of
the NMs.

The frequency shift of a single site oscillator induced by the nonlinearity is
$\delta_l = \beta |\psi_l|^{2}$. If instead a single mode is excited, its
frequency shift is given by $\delta_{\nu} = \beta |\phi_{\nu}|^2/
p_{\nu}$.

\subsection{Anharmonic oscillator lattice}

The Hamiltonian of the quartic Klein-Gordon lattice (KG)
\begin{equation}
\mathcal{H}_{K}= \sum_{l}  \frac{p_{l}^2}{2} +
\frac{\tilde{\epsilon}_{l}}{2} u_{l}^2 +
\frac{1}{4} u_{l}^{4}+\frac{1}{2W}(u_{l+1}-u_l)^2,
\label{RQKG}
\end{equation}
where $u_l$ and $p_l$ are respectively the generalized coordinates and
momenta, and $\tilde{\epsilon}_{l}$ are
chosen uniformly from the interval $\left[\frac{1}{2},\frac{3}{2}\right]$.
The equations of motion are $\ddot{u}_{l} = - \partial \mathcal{H}_{K}
/\partial u_{l}$ and yield
\begin{equation}
\ddot{u}_{l} = - \tilde{\epsilon}_{l}u_{l}
-u_{l}^{3} + \frac{1}{W} (u_{l+1}+u_{l-1}-2u_l)\;.
\label{KG-EOM}
\end{equation}
Equations (\ref{KG-EOM}) conserve the energy (\ref{RQKG}). They serve e.g. as
simple models for the dissipationless dynamics of anharmonic optical lattice
vibrations in molecular crystals \cite{aaovchinnikov}.  The energy of an
initial state $E \geq 0$ serves as a control parameter of nonlinearity similar
to $\beta$ for the DNLS case.

The coefficient $1/ (2W)$ in (\ref{RQKG}) was chosen so that the linear parts
of Hamiltonians (\ref{RDNLS}) and (\ref{RQKG}) would correspond to the same
eigenvalue problem. In practice, for $E \rightarrow 0$ (or by neglecting the
nonlinear term $u_l^4/4$) model (\ref{RQKG}) with $u_{l} = A_{l} \exp(i\omega
t)$ is reduced to the linear eigenvalue problem (\ref{EVequation}) with
$\lambda = W\omega^2-W -2$ and $\epsilon_l=W( \tilde{\epsilon}_{l}-1)$.  The
width of the squared frequency $\omega_{\nu}^2$ spectrum is $\Delta_{K}= 1+
\frac{4}{W}$ with $\omega_{\nu}^2 \in \left[ \frac{1}{2},\frac{3}{2} +
\frac{4}{W}\right] $.  Note that $\Delta_D = W \Delta_{K}$. As in the case of
DNLS, $W$ determines the disorder strength.

The spatial properties of the NMs are identical with those of
(\ref{EVequation}). In addition to the scale $\Delta_{K}$, the average spacing
of squared eigenfrequencies of NMs within the range of a localization volume
is $\overline{\Delta \omega^2} = \Delta_{K} / \overline{p_{\nu}} $.  The two
scales $ \overline{\Delta \omega^2} \leq \Delta_{K} $ determine the packet
evolution details in the presence of nonlinearity.

The squared frequency shift of a single site oscillator induced by the
nonlinearity is $\delta_{l} \approx (3 E_l)/(2 \tilde{\epsilon}_l)$, where
$E_l$ is the energy of the oscillator.  If instead a single mode is excited,
its frequency shift is given by $\delta_{\nu} \approx (3 E_{\nu})/(2 p_{\nu}
\omega^2_{\nu}) $ with $E_{\nu}$ being the energy of the mode.

For small amplitudes the equations of motion of the KG chain can be
approximately mapped onto a corresponding DNLS model
\cite{KG-DNLS-mapping}. In our notation, the mapping takes the following
form. For the KG model with given parameters $W$ and $E$, the corresponding
DNLS model (\ref{RDNLS}) with norm $S=1$, has a nonlinearity parameter
$\beta\approx 3WE$.
The norm density of the DNLS model corresponds to the normalized energy
density of the KG model.

\subsection{Computational methods}

We will present results on long time numerical simulations. We therefore first
discuss the methods and particularities of our computations. For both models,
we used symplectic integrators. These integration schemes replace the original
Hamiltonian by a slightly different one, which is integrated exactly.  The
smaller the time steps, the closer both Hamiltonians. Therefore, the computed
energy (or norm) of the original Hamiltonian function will fluctuate in time,
but not grow. The fluctuations are bounded, and are due to the fact, that the
actual Hamiltonian which is integrated, has slightly different energy.

Another possible source of errors is the roundoff procedure of the actual
processor, when performing operations with numbers. Sometimes it is referred
to as `computational noise' although it is exactly the opposite, i.~e.~purely
deterministic and reproducible. We will discuss the influence of roundoff
errors on our results in section \ref{sec.roundoff}.

The KG chain was integrated with the help of a symplectic integrator of order
$\mathrm{\cal{O}}(\tau^4)$ with respect to the integration time step $\tau$,
namely the SABA$_2$ integrator with corrector (SABA$_2$C), introduced in
\cite{LR01}.  A brief presentation of the integration scheme, as well as its
implementation for the particular case of the KG lattice (\ref{RQKG}) is given
in Appendix \ref{sec:saba2c}. The SABA$_2$C integration scheme proved to be
very efficient for long integrations (e.~g.~up to $10^{10}$ time units) of
lattices having typically $N=1000$ sites (see for example the right plots of
Fig.~\ref{fig_ss}), since it kept the required computational time to feasible
levels, preserving at the same time quite well the energy of the system. For
example, an integration time step $\tau=0.2$ usually kept the relative error
of the energy smaller than $10^{-4}$.

The DNLS chain was integrated with the help of the SBAB$_2$ integrator (see
Appendix \ref{sec:saba2c}), which introduces an error in energy conservation
of the order $\mathrm{\cal{O}}(\tau^2)$. The number of sites used in our
computations varied from $N=500$ to $N=2000$, in order to exclude finite size
effects in the evolution of the wave packets. For $\tau=0.1$ the
relative error of energy was usually kept smaller than $10^{-3}$. It is worth
mentioning that, although the SBAB$_2$ integrator and the commonly used
leap--frog integrator introduce errors of the same
order, the SBAB$_2$ scheme exhibits a better
performance since it requires less CPU time, keeping at the same time the
relative energy error to smaller values than the leap--frog scheme.

We order the NMs in space by increasing value of the center-of-norm coordinate
$X_{\nu}=\sum_l l A_{\nu,l}^2$.  We analyze normalized distributions $z_{\nu}
\geq 0$ using the second moment $m_2= \sum_{\nu}
(\nu-\bar{\nu})^2 z_{\nu}$, which quantifies the wave packet's degree of
spreading and the participation number $P=1 / \sum_{\nu} z_{\nu}^2$, which
measures the number of the strongest excited sites in $z_{\nu}$.  Here
$\bar{{\nu}} = \sum_{\nu} \nu z_{\nu}$.  For DNLS we follow norm density
distributions $z_{\nu}\equiv |\phi_{\nu}|^2/\sum_{\mu} |\phi_{\mu}|^2$.  For
KG we follow normalized energy density distributions $z_{\nu}\equiv
E_{\nu}/\sum_{\mu} E_{\mu}$ with $E_{\nu} =
\dot{A}^2_{\nu}/2+\omega^2_{\nu}A_{\nu}^2/2$, where $A_{\nu}$ is the amplitude
of the $\nu$th NM and $\omega^2_\nu=1+(\lambda_{\nu}+2)/W$.

\section{Wave packet evolution}
\label{sec.regimes}

Below we will mainly use the DNLS case for theoretical considerations, and
also discuss crucial points to be taken into account, when considering the KG
case. We will present numerical results for both models.

We first consider a wave packet at $t=0$ which is compact either in real
space, or in normal mode space. Compactness in real space implies a single
site excitation $\psi_{l} = \delta_{l,l_0}$ with the choice $\epsilon_{l_0}=0$
for the DNLS model. For the KG model we set $p_l=0$, $u_l=c \,\delta_{l,l_0}$
with $\tilde{\epsilon}_{l_0}=1$ and $c$ being a constant which defines the
initial energy $E$.  Compactness in normal mode space instead implies a single
mode excitation
$\phi_{\nu} = \delta_{\nu,\nu_0}$ with $\lambda_{\nu_0} \approx 0$ for the
DNLS model, while in the case of the KG system we have $A_{\nu}=c
\,\delta_{\nu,\nu_0}$, $\dot{A}_\nu=0$, with $\omega_{\nu_0}^2 \approx
1+(2/W)$, i.~e.~$\omega_{\nu_0}^2$ is located in the middle of the frequency
spectrum. Again the constant $c$ defines the initial energy of the wave
packet.  We will later also consider finite size initial distributions of
width $L$.

\subsection{Expected regimes}
\label{Sec:reg}

Let us consider a single site initial excitation with a corresponding
nonlinear frequency shift $\delta_l$. We compare this frequency shift with the
two scales set by the linear equations: the average spacing $\overline{\Delta
\lambda}$ (which corresponds to $\overline{\Delta \lambda}_D$ for DNLS and to
$\overline{\Delta \omega^2}$ for KG) and the spectrum width $\Delta$ (with
$\Delta$ denoting $\Delta_D$ for DNLS and $\Delta_K$ for KG).  We expect three
qualitatively different dynamical regimes: I) $\delta_l < \overline{\Delta
\lambda}$; II) $\overline{\Delta \lambda} < \delta_l < \Delta$; III) $\Delta <
\delta_l$. In case I the local frequency shift is less than the average
spacing between interacting modes, therefore no initial resonance overlap of
them is expected, and the dynamics may (at least for long times) evolve as in
the linear case ($\beta=0$ for DNLS and $E \rightarrow 0$ for KG).  In case II
resonance overlap may happen immediately, and the packet should evolve
differently.  For case III the frequency shift exceeds the spectrum width,
therefore some renormalized frequencies of NMs (or sites) may be tuned out of
resonance with the NM spectrum, leading to selftrapping.  The above
definitions are highly qualitative, since localized initial conditions are
subject to strong fluctuations.

If we instead consider a single mode initial excitation, we have to replace
$\delta_l$ by $\delta_{\nu}$ in the above argumentation. For both the DNLS
 and the KG model, it follows $\delta_l \sim
p_{\nu} \delta_{\nu}$.  The mean NM participation number (the localization
volume) $\overline{p_{\nu}} >1$ depends on the disorder strength $W$.

\begin{figure}
\centerline{
\begin{tabular}{cc} \hspace{0.3cm}
\includegraphics[scale=0.22]{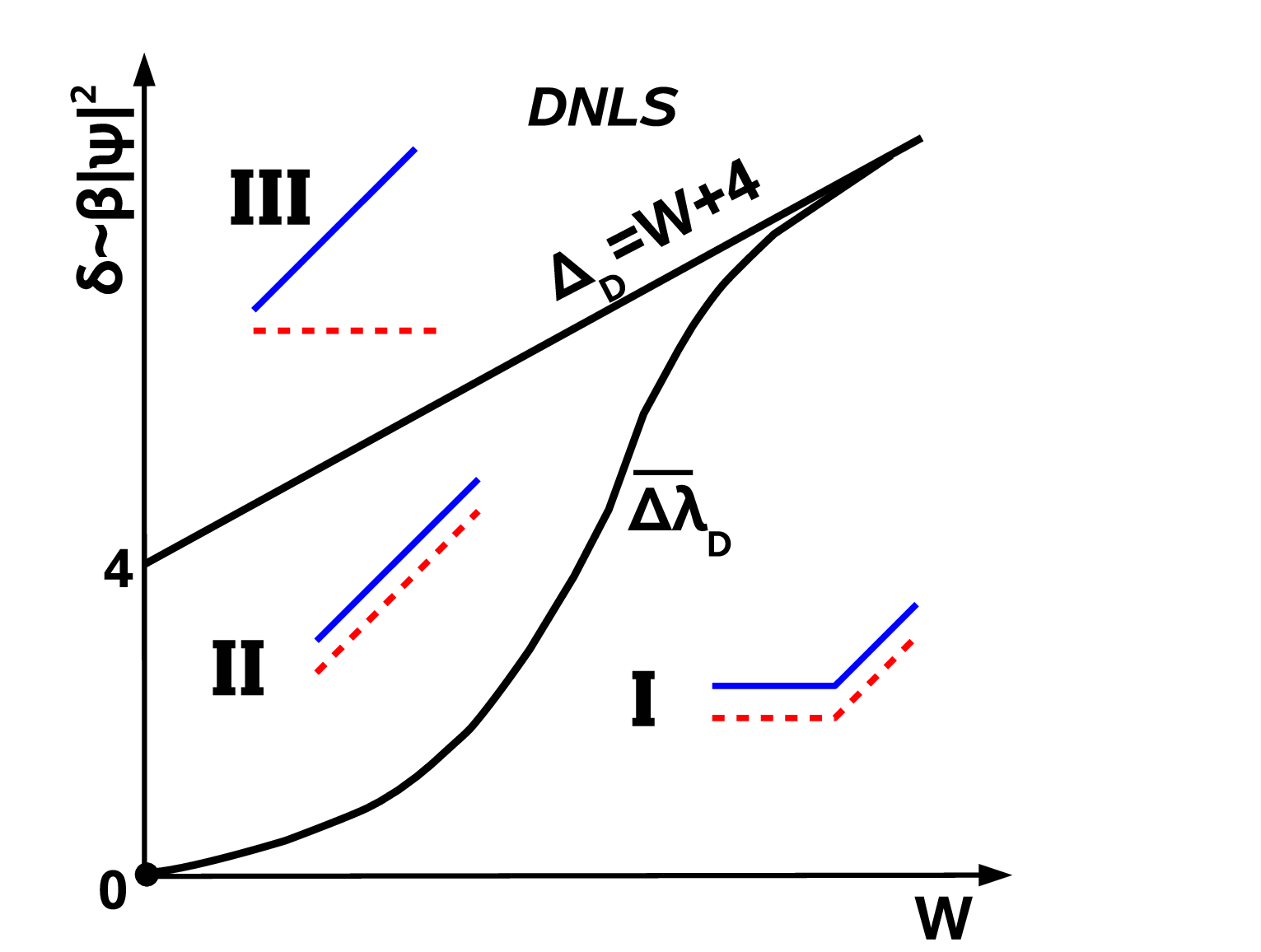} \hspace{-1.5cm}
\includegraphics[scale=0.22]{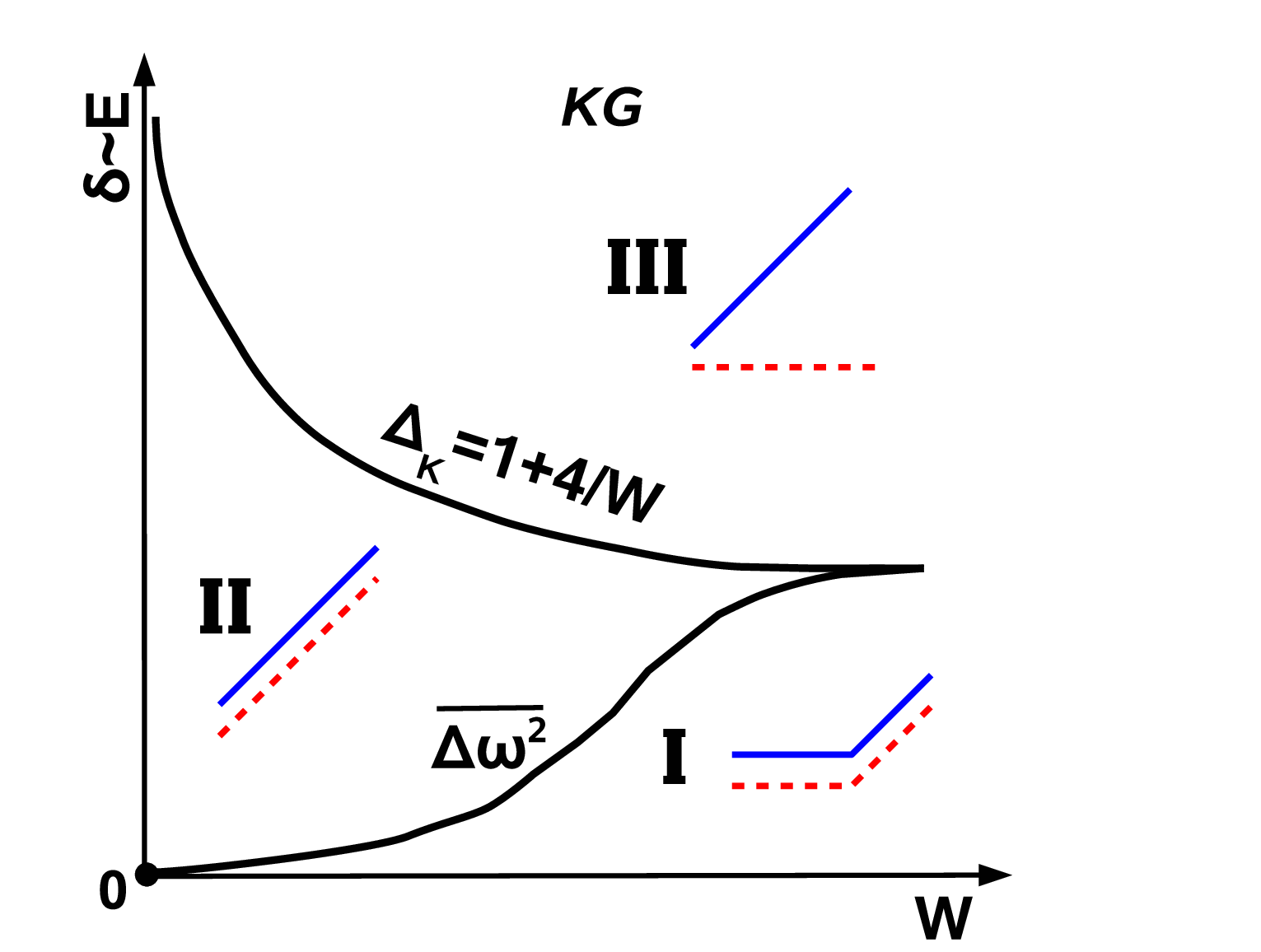}  \end{tabular}}
\caption{(color online) Schematic representations of the three different
  regimes of spreading for the DNLS (left graph) and the KG model (right
  graph), in the parameter space of disorder strength $W$ and of the nonlinear
  frequency shift $\delta$ at initial time $t=0$. For each regime the
  dependence of $\log m_2$ (blue solid curves) and of $\log P$ (red dashed
  curves) versus $\log t$ are shown schematically (see section
  \ref{sec.numerics} for details).  }
\label{fig_param}
\end{figure}

If an initial excitation of the DNLS model is characterized by some
exponentially localized (not necessarily compact) distribution $\psi_l$ with
$S=1$, the nonlinear frequency shift may be roughly estimated as $\delta \sim
\beta |\psi|^2$, where the maximum norm density $|\psi|^2 =
\sup_l|\psi_l|^2$. The left graph of Fig.\ref{fig_param} shows the location
of the three different regimes in the plane of the control parameters,
i.~e.~the frequency shift $\delta$ and the disorder strength $W$.
Note that $\overline{\Delta \lambda} \propto W^3$ for $W \ll 1$ \cite{MIRLIN},
and the intermediate regime II disappears around $W \approx20$, where the
participation number of a NM becomes of the order of one, and the NMs become
almost single site solutions. Similarly, for the KG model we have the
estimation $\delta \sim E$ and the corresponding parameter space of the three
different regimes is shown in the right graph of Fig.\ref{fig_param}.

\subsection{The selftrapping theorem}

Regime III is also captured by a theorem presented in \cite{kkfa08}, which
proves, that for $\beta > \Delta$ (for the DNLS case) the single site
excitation can not uniformly spread over the entire (infinite) lattice.
Indeed, with the notations
\begin{eqnarray}
\mathcal{H}_{D}= \mathcal{H}_{NL}+\mathcal{H}_{L} \;,
\label{stt-1} \\
\mathcal{H}_{L} = \sum_{l} \epsilon_{l}
|\psi_{l}|^2- (\psi_{l+1}\psi_l^*  +\psi_{l+1}^* \psi_l)\;,
\label{stt-2} \\
\mathcal{H}_{NL} = \sum_{l} \frac{\beta}{2} |\psi_{l}|^{4} \equiv
\frac{\beta}{2} P_r^{-1}\;,
\label{stt-3}
\end{eqnarray}
where $P_r$ is the participation number in real space, the single site
excitation at time $t=0$ yields
\begin{equation}
\mathcal{H}_{L}(t=0) = 0\;,\; \mathcal{H}_{NL}(t=0) = \frac{\beta}{2}\;.
\label{sst-4}
\end{equation}
Due to norm conservation $S=1$ at all times, the harmonic energy part
$\mathcal{H}_{L}$ is bounded from above and below \cite{kkfa08}:
\begin{equation}
-2 -\frac{W}{2} \leq \mathcal{H}_{L} \leq 2 + \frac{W}{2}\;.
\label{sst-5}
\end{equation}
Due to energy conservation, for all times the anharmonic energy part
$\mathcal{H}_{NL}$ can therefore not become smaller than
\begin{equation}
\mathcal{H}_{NL}(t) \geq \frac{\beta}{2}-2-\frac{W}{2}\;.
\label{sst-6}
\end{equation}
It follows with (\ref{stt-3}), that the participation number is bounded from
above by a finite number, which diverges for $\beta = \Delta$:
\begin{equation}
P_r(t) \leq \frac{\beta}{\beta - \Delta}\;{\rm if}\; \beta \geq \Delta\;.
\label{sst-7}
\end{equation}
Moreover, since $P_r^{-1}=\sum_l |\psi_l|^4 < \sup_l |\psi_l|^2$ \cite{kkfa08},
we conclude that
\begin{equation}
\sup_l |\psi_l|^2 (t) > \frac{\beta-\Delta}{\beta}\;.
\label{sst-8}
\end{equation}
Therefore, at least a part of the wave packet will not spread, and stay
localized, although the theorem does not prove that the location of that
inhomogeneity is constant in time. The norm of the part of the wave packet,
which can spread uniformly over the entire system, is bounded from above by
$S_{\infty} \leq \Delta / \beta$.

\subsection{Numerical results}
\label{sec.numerics}

\begin{figure}
\includegraphics[angle=0,width=0.99\columnwidth]{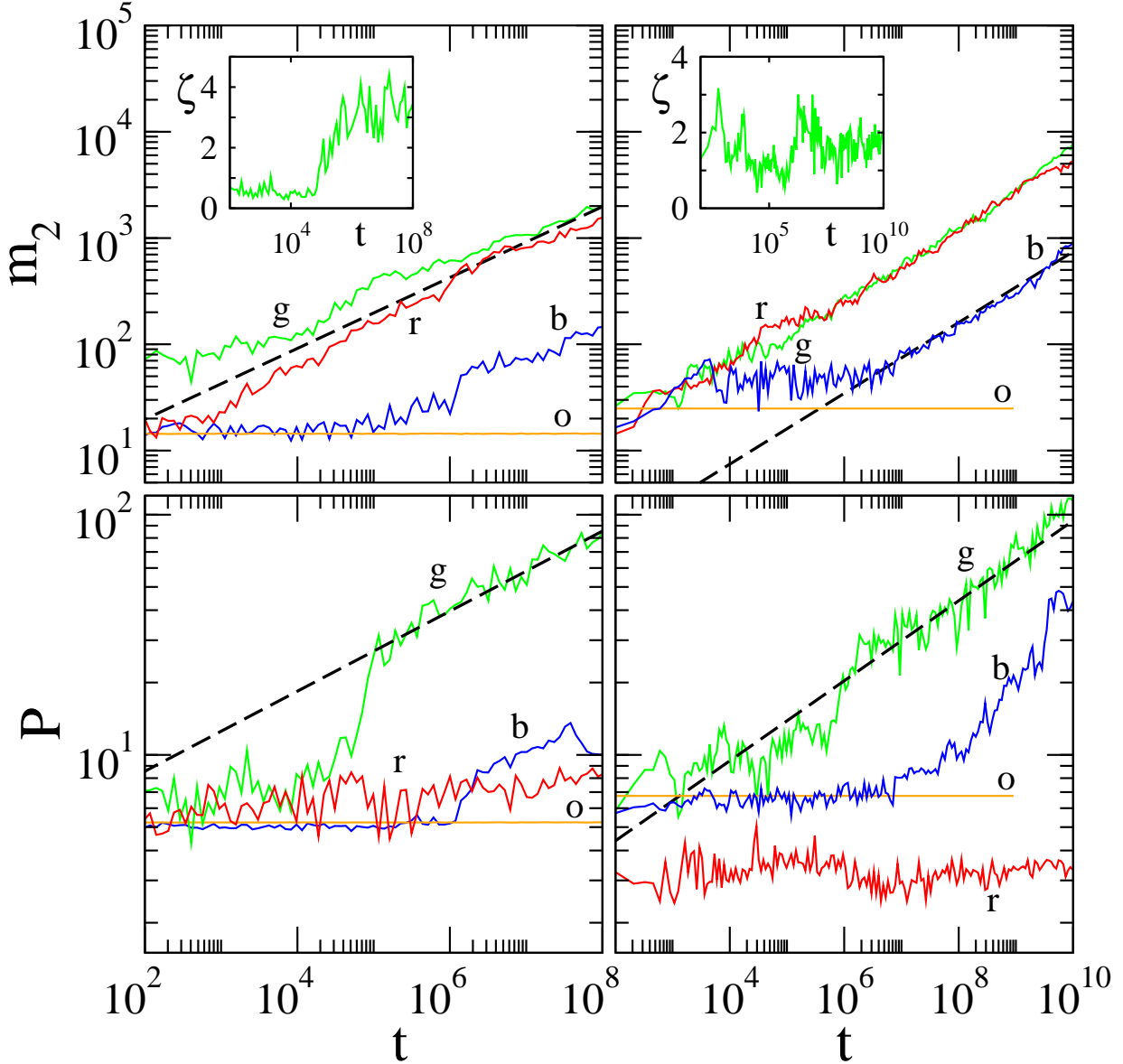}
\caption{(color online) Single site excitations.  $m_2$ and $P$ versus time in
log--log plots.  Left plots: DNLS with $W=4$, $\beta=0,0.1,1,4.5$ [(o),
orange; (b), blue; (g) green; (r) red].  Right plots: KG with $W=4$ and
initial energy $E=0.05,0.4,1.5$ [(b) blue; (g) green; (r) red]. The orange
curves (o) correspond to the solution of the linear equations of motion, where
the term $u_l^3$ in (\ref{KG-EOM}) was absent.  The disorder realization is
kept unchanged for each of the models.  Dashed straight lines guide the eye
for exponents 1/3 ($m_2$) and 1/6 ($P$) respectively. Insets: the compactness
index $\zeta$ as a function of time in linear--log plots for $\beta=1$ (DNLS)
and $E=0.4$ (KG). }
\label{fig_ss}
\end{figure}

We first show results for single site excitations \cite{fks08}.  We
systematically studied the evolution of wave packets for lattices (\ref{RDNLS})
and (\ref{RQKG}).  The scenario described in section \ref{Sec:reg} was
observed very clearly. Representative examples are shown in Fig.\ref{fig_ss}.
Regime III yields selftrapping (see also Figs.~1, 3 in \cite{kkfa08}),
therefore $P$ does not grow significantly, while the second moment $m_2\sim
t^{\alpha}$ with $\alpha \approx 1/3$ (red curves).  Thus a part of the
excitation stays highly localized \cite{kkfa08}, while another part
delocalizes. Regime II yields subdiffusive spreading with $m_2\sim t^{\alpha}$
and $P \sim t^{\alpha/2}$ \cite{Mol98,PS08} (green curves).  Regime I shows
Anderson localization up to some time  $\tau_d$ which increases with
decreasing nonlinearity. For $t < \tau_d$ both $m_2$ and $P$ are not changing.
However for $t > \tau_d$ a detrapping takes place, and the packet starts to
grow with characteristics as in regime II (blue curves).  The
simulation of the equations of motion in the absence of nonlinear terms (orange
curves), demonstrates the appearance of Anderson localization.

The second moment $m_2$ is sensitive to the spreading distance of the tails of
a distribution, while the participation number $P$ is a measure of the
inhomogeneity of the distribution, being insensitive to any spatial
correlations. Thus, $P$ and $m_2$ can be used to quantify the
sparseness of a wave packet. To this end, we introduce as a measure of the
compactness of a wave packet the compactness index
\begin{equation}
\zeta=\frac{P^2}{m_2}.
\label{eq:ci}
\end{equation}

Let us consider a wave packet of $K$ sites ($K \gg 1$). In the case where all
the $K$ sites are equally excited the compactness index is given by
$\zeta=12$. In the case of a symmetric wave packet formed by a sequence of an
excited site followed by a nonexcited one, where all the $K/2$ excited sites
have the same amplitude, $\zeta=3$. Distributions with larger gaps between the
equally excited isolated sites attain a compactness index $\zeta<3$. For the
extreme case of a sparse wave packet formed by two equally excited sites
located at the two edges of the packet, i.~e.~when only sites 1 and $K$ $(K
\gg 1 )$ are excited to an amplitude 1/2, the compactness index is
$\zeta=16/K^2$. So, smaller values of $\zeta$ correspond to more sparse wave
packets.

\begin{figure}
\includegraphics[angle=0,width=0.99\columnwidth]{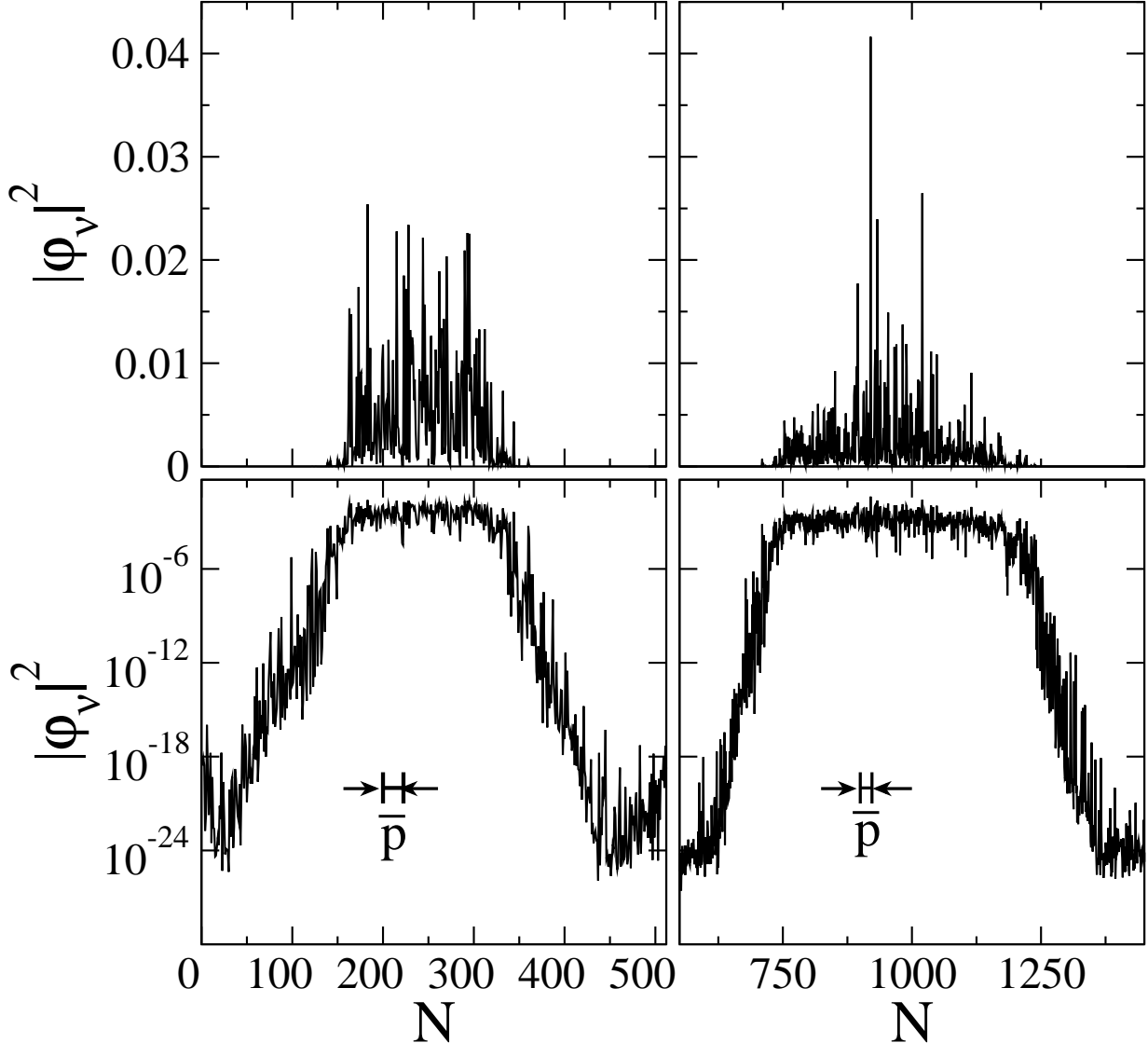}
\caption{Norm density distributions in the NM space at time
$t=10^8$ for the initial excitations in the regime II of the DNLS model shown
in the left plots of Figs.~\ref{fig_ss} and \ref{fig_sm}. Left plots: single
site excitation for $W=4$ and $\beta=1$. Right plots: single mode
excitation for $W=4$ and $\beta=5$.  $|\phi_\nu|^2$ is plotted in
linear (logarithmic) scale in the upper (lower) plots. The maximal mean value
of the localization volume of the NMs $\overline{p}\approx22$ (shown
schematically in the lower plots) is much smaller than the length over
which the wave packets have spread.  }
\label{fig_dis}
\end{figure}

We expect that $\zeta$ in regime I will remain constant for $t< \tau_d$
and will behave as in the case of regime II for latter times. In regime II
$\zeta$ would either be constant or decay in time, while in regime III
it should decay since $P$ remains practically constant. The time evolution of
$\zeta$ for excitations in regime II is shown in the insets of
Fig.~\ref{fig_ss}. As one can see the compactness index oscillates around some
constant nonzero value both for the DNLS and the KG models. This means that
the wave packet spreads but does not become more
sparse. For the particular cases of Fig.~\ref{fig_ss} the compactness index
attains the values $\zeta=3.5$ for the DNLS model at $t=10^8$ and
$\zeta=1.7$ for the KG chain at $t=10^{10}$. The corresponding wave
packet of the DNLS model is shown in the left plots of Fig.~\ref{fig_dis}.

Partial nonlinear localization in regime III is explained by selftrapping
\cite{kkfa08}. It is due to tuning frequencies of excitations out of resonance
with the NM spectrum, takes place irrespective of the presence of disorder and
is related to the presence of exact $t$-periodic spatially localized states
(also coined discrete breathers) for ordered \cite{DB} and disordered systems
\cite{DDB} (in the latter case also $t$-quasiperiodic states exist). These
exact solutions act as trapping centers.

\begin{figure}
\includegraphics[angle=0,width=0.99\columnwidth]{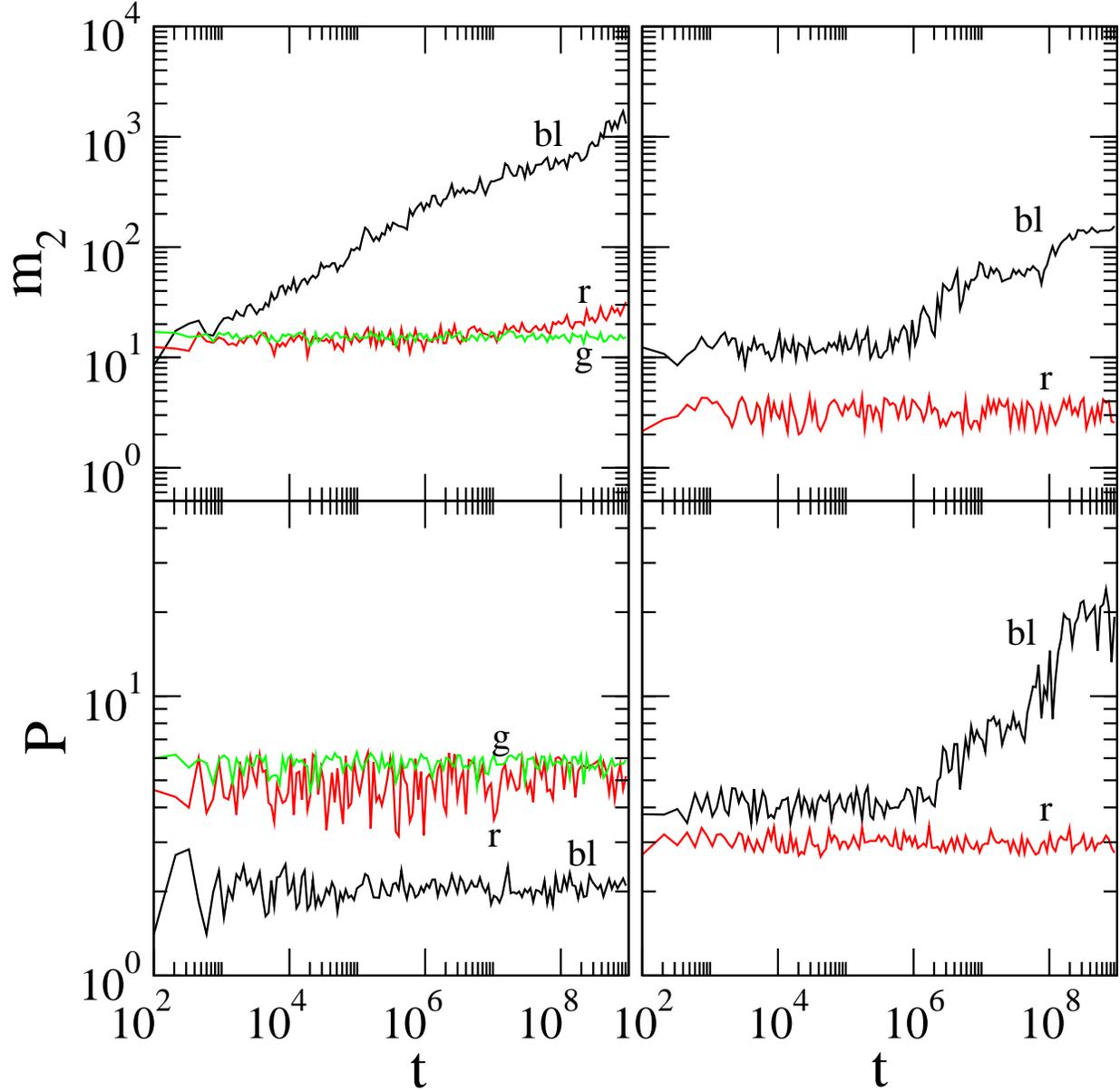}
\caption{(color online) Single site excitations for the same disorder
realization of the KG model. $m_2$ and $P$ versus time in log--log plots. Left
panels: plots for $W=4$ and initial energy $E=3.225, 4, 10$ [(bl) black; (r)
red; (g) green]. Right panels: Plots for $E=0.05$ and $W=6, 7$ [(bl) black;
(r) red].}
\label{fig_EW}
\end{figure}

Note that for large nonlinearities ($\beta \gg \Delta$ for DNLS or large
energy values $E$ of the KG model) almost the whole excitation is
selftrapped. This behavior can be seen in the left plots of
Fig.~\ref{fig_EW}, where the time evolution of $m_2$ and $P$ for different
values of the energy E of the KG chain is shown.
The value of $W$ is kept to $W=4$ as in the cases presented in the right
plots of Fig.~\ref{fig_ss}. As the energy increases  the
portion of the wave packet that stays selftrapped increases with respect to
the part that diffuses. Thus, we observe a change in the evolution of $m_2$
from subdiffusive increase to practical constancy. On the other
hand, $P$ is not affected as it continues to fluctuate around some constant
value.

Anderson localization on finite times in regime I is observed on potentially
large time scales $\tau_d$, and as in III, regular states act as trapping
centers \cite{DDB}. For $ t > \tau_d$, the wave packet trajectory finally
departs away from the vicinity of regular orbits, with subsequent
spreading. Increasing the value of $W$ results to small localization lengths
of NMs and thus, Anderson localization will persist for extremely long time
intervals. Since our numerical computations are limited in time, we are not
able to observe the detrapping phase of the evolution when $W$ increases
significantly. This behavior can be seen in the right plots of
Fig.~\ref{fig_EW} where we consider initial single site excitations which, for
$W=4$ (see right plots of Fig.~\ref{fig_ss}) belong to regime I. In these
plots we observe a direct transition from regime I to practical constancy of
$m_2$ and $P$ as $W$ increases, at least up to the final integration time
used.

\begin{figure}
\includegraphics[angle=0,width=0.99\columnwidth]{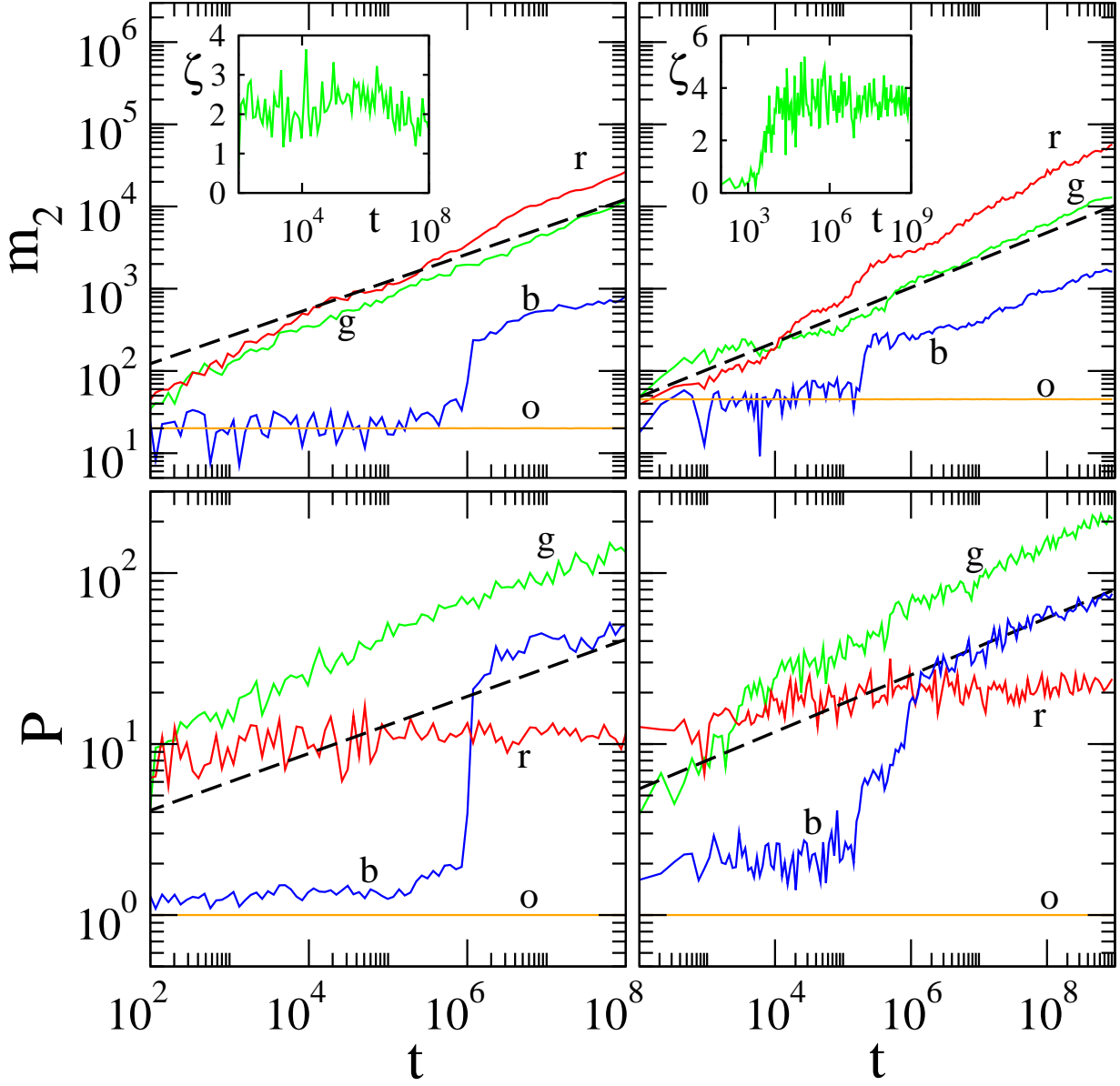}
\caption{(color online) Single mode excitations.  $m_2$ and $P$ versus time in
log--log plots.  Left plots: DNLS with $W=4$, $\beta=0,0.6,5,30$ [(o) orange;
(b) blue; (g) green; (r) red].  Right plots: KG with $W=4$ and initial energy
$E=0.17,1.1,13.4$ [(b) blue; (g) green; (r) red].  The orange curves (o)
correspond to the solution of the linear equations of motion, where the term
$u_l^3$ in (\ref{KG-EOM}) was absent.  The disorder realization is kept
unchanged for each of the models.  Dashed straight lines guide the eye for
exponents 1/3 ($m_2$) and 1/6 ($P$) respectively. Insets: the compactness
index $\zeta$ as a function of time in linear--log plots for $\beta=5$ (DNLS)
and $E=1.1$ (KG).}
\label{fig_sm}
\end{figure}

For single mode excitations we find a similar outcome, but with rescaled
critical values for the nonlinearity strength which separate the different
regimes.  Examples of the three different regimes are shown in
Fig.\ref{fig_sm}.
As in the case of single site excitations presented in Fig.~\ref{fig_ss}, the
compactness index $\zeta$ plotted in the insets if Fig.~\ref{fig_sm}
remains practically constant for excitations in regime II, attaining the
values $\zeta=1.5 $ at $t=10^8$ for the DNLS model and $\zeta=3.3
$ at $t=10^9$ for the KG chain.  The final norm density distribution for the
DNLS model is plotted in the right plots of Fig.~\ref{fig_dis}. The average
value $\overline{\zeta}$ of the compactness index over 20 realizations
at $t=10^8$ for the DNLS model with $W=4$ and $\beta=5$ was found to be
$\overline{\zeta}=2.95 \pm 0.39$.

\subsection{Spreading}
\label{sec:spread}

\begin{figure}
\includegraphics[angle=0,width=0.99\columnwidth]{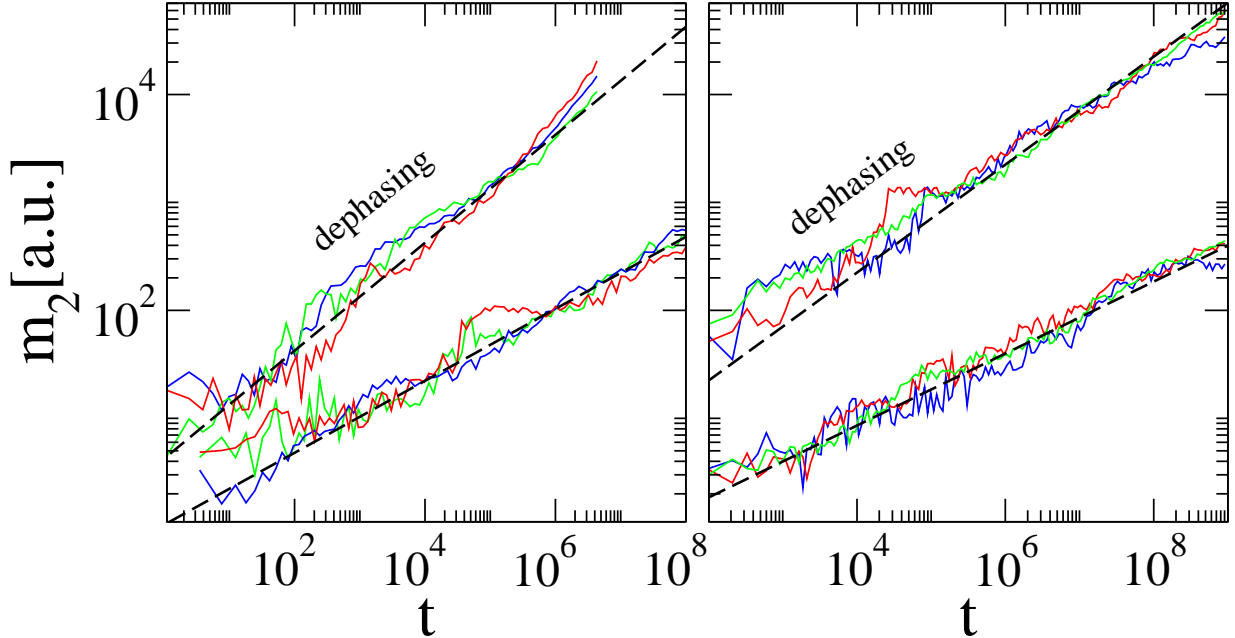}
\caption{(color online) Single site excitations.  $m_2$ (in arbitrary units)
versus time in log--log plots in regime II and different values of $W$.  Lower
set of curves: plain integration (without dephasing); upper set of curves:
integration with dephasing of NMs (see section \ref{sec:dephase}).  Dashed
straight lines with exponents 1/3 (no dephasing) and 1/2 (dephasing) guide the
eye.  Left plot: DNLS, $W=4$, $\beta=3$ (blue); $W=7$, $\beta=4$ (green);
$W=10$, $\beta=6$ (red).  Right plot: KG, $W=10$, $E=0.25$ (blue) , $W=7$,
$E=0.3$ (red) , $W=4$, $E=0.4$ (green).  The curves are shifted vertically in
order to give maximum overlap within each group.}
\label{fig_dephase}
\end{figure}

The subdiffusive spreading takes place in regime I for $t>\tau_d$, in regime
II, and for a part of the wave packet also in regime III.  For single site
excitations the exponent $\alpha$ does not appear to depend on $\beta$ in the
case of the DNLS model or on the value of $E$ in the case of KG. In
Fig.\ref{fig_dephase}
we show results for $m_2(t)$ in  regime II for different values
of the disorder strength $W$.  Again we find no visible dependence of the
exponent $\alpha$ on $W$.  Therefore the subdiffusive spreading is rather
universal and the parameters $\beta$ (or $E$) and $W$ are only affecting the
prefactor.  Excluding selftrapping, any nonzero
nonlinearity  will completely delocalize the wave packet and destroy Anderson
localization.  We performed fittings by analyzing 20 runs in regime II with
different disorder realizations. For each realization we fitted the exponent
$\alpha$, and then averaged over all computational measurements.  We find
$\alpha = 0.33 \pm 0.02$ for DNLS, and $\alpha = 0.33 \pm 0.05$ for KG.
Therefore, the predicted universal exponent $\alpha=1/3$ \cite{fks08} appears
to explain the data.

On the other hand, in the case of single mode excitations the numerically
computed values of the exponent $\alpha$ seem to be slightly larger than
$\alpha=1/3$, as can be also seen from the results of Fig.~\ref{fig_sm}. In
particular, $m_2$ in regimes II and III of the DNLS model and in regime III of
the KG model increases slightly faster than $\propto t^{1/3}$, which is
represented by the dashed lines in the upper plots of Fig.~\ref{fig_sm}. In
addition, the value of the exponent seems to slightly vary with respect to the
nonlinearity parameter $\beta$ for DNLS and $E$ for KG. The reason of the
slightly different behavior between single site and single mode excitations is
still an open issue.

\subsection{Detrapping}

In the intermediate regime II the wave packet starts to spread almost from
scratch.  We do not observe any saturation and crossover into localization on
later times.  Let us assume that the wave packet spreads without
limitations. The initial nonlinear frequency shift $\delta_l$ was larger than
the average level spacing in a localization volume $\overline{\Delta
\lambda}$.  However, $\delta_l$ will become smaller than $\overline{\Delta
\lambda}$ at some later time, since $\sup_l|\psi_l|^2$ ($\sup_l E_l$ for KG)
decreases in time as the wave packet spreads.  Therefore, there will be a large
but finite time $t_d$, at which we cross over from the intermediate regime II
into the weak nonlinearity regime I.  The arresting of the wave packet up to a
time $\tau_d$ in the weak nonlinearity regime I can be explained by a
correspondingly large spreading time scale $\tau_d$. For $t <\tau_d$ no
spreading is observed when monitoring the second moment $m_2$, with subsequent
spreading observed on larger time scales $t > \tau_d$.

\begin{figure}
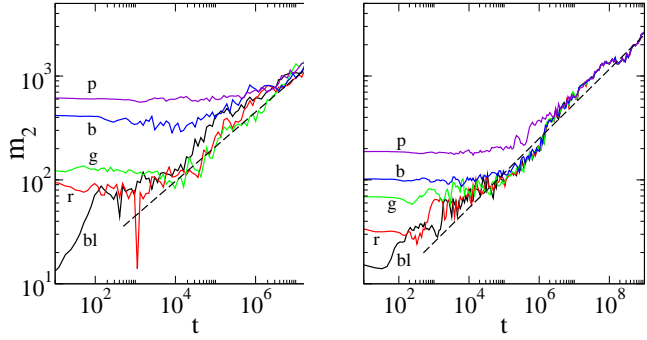

\centerline{
\begin{tabular}{cc}
\includegraphics[scale=0.247]{ff_07an.eps}
\includegraphics[scale=0.247]{ff_07bn.eps}\end{tabular}}
\caption{(color online) Evolution of $m_2$ versus time in log--log
plots. Single site excitations in the intermediate regime II for the DNLS
(left plot) and the KG model (right plot) correspond to black curves (bl).
The wave packets after $t_d=10^3$, $10^4$, $10^5$, $10^6$ time units (t.~u.)
[(r) red; (g) green; (b) blue; (p) purple] are registered and relaunched as
initial distributions (colored curves). The dashed straight lines correspond
to functions $\propto t^{1/3}$.}
\label{fig_td}
\end{figure}

We test the above conclusions by the following simple scheme. We start a
single site excitation in the intermediate regime II, measure the distribution
at some time $t_d$, and relaunch the distribution as an initial condition at
time $t=0$. The results are shown in Fig.\ref{fig_td}.  We find that the
relaunched runs yield a second moment $m_2$ which appears to be constant up to
the time $\tau_d \approx t_d$ with a subsequent spreading, similar to the
previously obtained detrapping in  regime I.

For a specific value of the nonlinearity $\beta$ of the DNLS model let each NM
in the packet after some spreading to have norm $|\phi_{\nu}|^2 \sim n \ll 1$
with $n$ denoting the average norm density of the excited NMs (in the case of
the KG model $n$ corresponds to the average energy density of the excited
NMs). The packet size is then $1/n \gg \overline{p}$, with
$\overline{p}=\max_{\nu} \overline{p_{\nu}}$, and the second moment $m_2 \sim
1/n^2$.  Let us assume that the second moment  grows as $m_2
\sim t^{1/3}$.  Let us also assume, that at any time the spreading is due to
some diffusion process, and is characterized by some momentary diffusion rate
$D(t)$ such that $m_2 =D(t) t$.  Then it follows that $D(t) \sim t^{-2/3}$ and
finally $D \sim n^4$.  Such a result has to be the outcome of the action of
the nonlinear terms (which always contain products $\beta n$). A diffusion
rate is equal to an inverse characteristic time scale, and therefore we
conjecture
\begin{equation}
D = \tau_d^{-1} \sim \beta^4 n^4\;.
\label{diffcoef}
\end{equation}
There are two ways of modifying $D$. We can either spread our initial
excitation over some number of sites $L$, therefore varying $n$. Alternatively
we can fix the shape of the initial excitation, and vary $\beta$.

\begin{figure}
\includegraphics[angle=0,width=0.99\columnwidth]{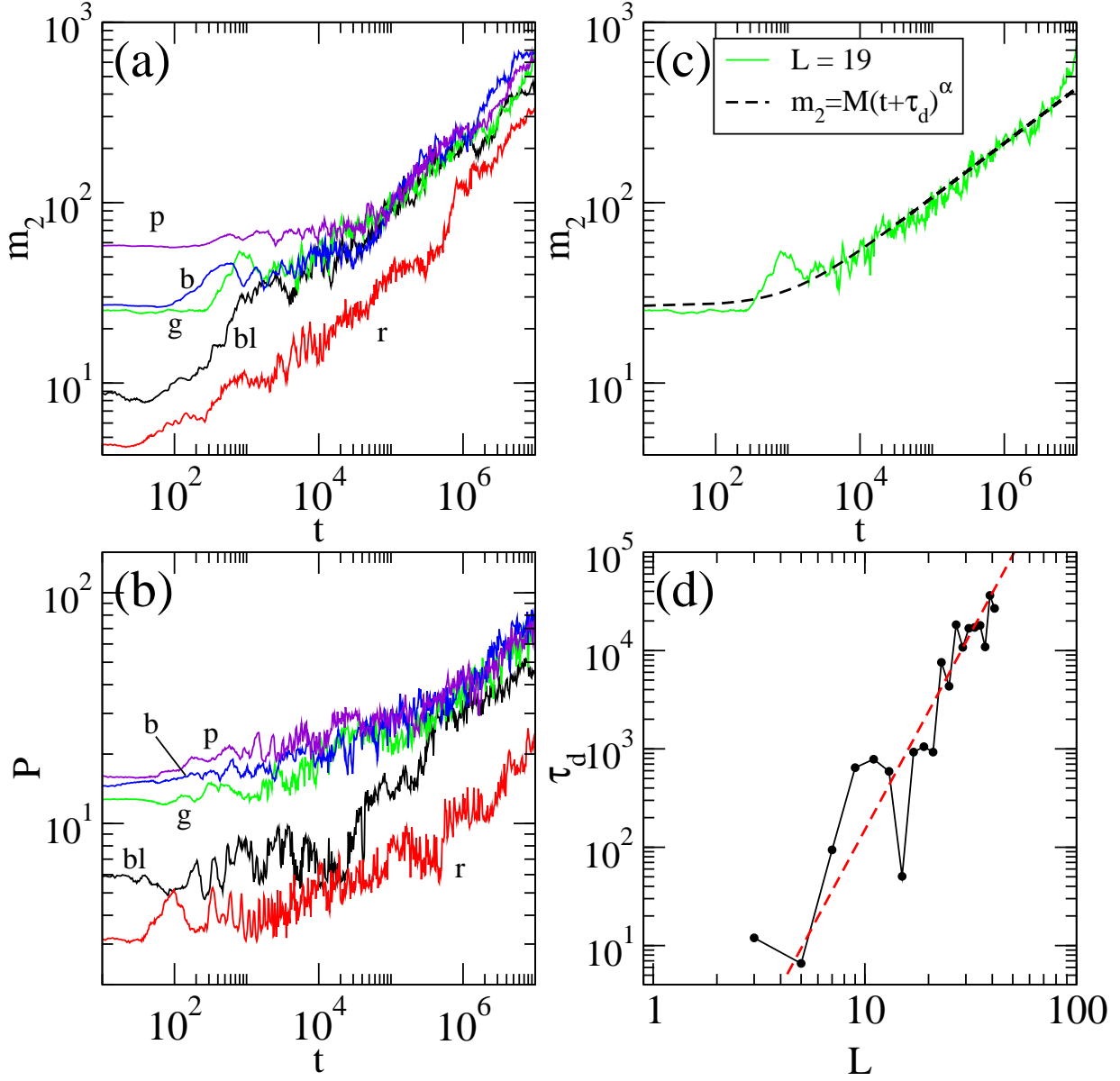}
\caption{(color online) Nonlocal excitations of the KG chain corresponding to
initial homogeneous distributions of energy $E=0.4$ over $L$ neighboring
sites. (a) $m_2$ and (b) $P$ versus time in log--log plots for $L=$1, 9, 19,
29 and 39 sites [(bl) black; (r) red; (g) green; (b) blue; (p) purple]. (c)
Fitting of the time evolution of $m_2$ for $L=19$ with a curve of the form
(\ref{eq:fit}) for $M=3.25$, $\tau_d=1052$ and $\alpha=0.303$. (d) The
dependence of the detrapping time $\tau_d$ on the number $L$ of initially
excited sites in log--log scale. The dashed straight line corresponds to a
function $\propto L^4$.}
\label{fig_L}
\end{figure}

In order to test the validity of Eq.~(\ref{diffcoef}) for a fixed value of
nonlinearity we  considered a single site
excitation in the intermediate regime II for the KG model with total energy
$E=0.4$, so that $m_2$ and $P$ start to grow from the beginning (black curves
in Figs.~\ref{fig_L} (a) and (b) respectively).
We also followed the time evolution of wave packets having as initial condition
a homogeneous distribution of the energy $E=0.4$ among $L$ neighboring
sites. In particular, we considered initial distributions with $u_l=0$ and
$p_l=0$ except for the central $L$ sites whose initial momenta were set to
$\pm \sqrt{2E/L}$, with the sign changing randomly from site to site. We
performed simulations with $L$ ranging from 1 up to 41. The time evolution of
$m_2$ and $P$ for some of these cases is shown in Figs.~\ref{fig_L}(a) and (b)
respectively. In accordance to the results presented in Fig.~\ref{fig_td} we
observe that, distributing the energy of a single site excitation belonging to
regime II over more sites results in a time dependence of $m_2$ and $P$
similar to regime I, i.~e.~both quantities start to increase after some
transient detrapping time $\tau_d$.

The behavior of the second moment $m_2(t)$ can be
modeled by a function of the form
\begin{equation}
m_2(t)=M(t+\tau_d)^{\alpha},
\label{eq:fit}
\end{equation}
where $M$ is a constant related to the value of the second moment of the
initial distribution $m_2(0)=M \tau_d^{\alpha}$. Eq.~(\ref{eq:fit}) gives a
power law dependence of $m_2$ on $t$ for $t \gg \tau_d$ and a slow time
dependence of $m_2$ for $t\ll \tau_d$. Thus, it can be used to describe the
behavior of $m_2$ for $L>1$. Fitting the numerical data obtained for
different values of $L$ by Eq.~(\ref{eq:fit}) (see Fig.~\ref{fig_L}(c) for
such an example) we can determine the dependence of $\tau_d$ on $L$
(Fig.~\ref{fig_L}(d)). Since $L \sim n^{-1}$ from (\ref{diffcoef}) we conclude
that $\tau_d \sim L^4$. As we can see from Fig.~\ref{fig_L}(d) the numerically
obtained results are in good agreement with this assumption.

To test the dependence of $D$ on $\beta$, we studied the weak nonlinearity
regime I for the DNLS model with $W=4$. We launched single site excitations for
10 realizations for $\beta=0.1$ and $\beta=0.2$. We estimated the detrapping
times $\tau_d$ on logarithmic scale for each run, and averaged over each group
of realizations.  As a result we obtain $\langle \log_{10} \tau_d \rangle =5$
for $\beta=0.2$, and $\langle \log_{10} \tau_d \rangle =6.9$ for $\beta=0.1$
(with $\langle \cdots \rangle$ denoting the mean value over the realizations),
and their difference is then 1.9.  According to (\ref{diffcoef}), the
difference should be $1.2$ which is in relatively good
agreement with the numerically estimated value.

\subsection{Numerical accuracy and roundoff errors}
\label{sec.roundoff}

We performed several tests in order to ensure that our results are not
generated by inaccurate computations. First we varied the size of the system
and found no dependence of the results on it.  Therefore we exclude finite
size effects.

Second we varied the time steps of the symplectic integration schemes by
orders of magnitudes. Again we found no visible change in the detrapping
times, or in the spreading characteristics. We also used different integration
schemes, and even nonsymplectic ones (8th order Runge-Kutta). No changes were
obtained either.  Therefore we exclude effects due to discretization of time.

Finally we studied the influence of computational roundoff errors. The above
observation, that the variation of time steps does not change the key results,
implicitly tells that roundoff errors can be excluded as well. Indeed,
changing the time steps, we change the number of operations to be performed on
a given interval of integration. Therefore we change the number of roundoff
operations.

In addition, we decided to perform further tests with respect to the roundoff
error issue. These tests are inspired by the following consideration. Floating
point numbers are characterized by the number of digits $a$ after the comma
which are kept during computations. All presented data were obtained with
double precision, where $a=16$.  The detrapping and spreading can be only due
to the cubic nonlinear terms in the equations of motion.  These terms are
added to linear terms, when calculating the rhs of (\ref{RDNLS-EOM}) and
(\ref{KG-EOM}).  Therefore, when for example in the case of the DNLS model
$\sup_l|\psi_l|^2 < 10^{-8}$, the nonlinear terms become of the order of the
roundoff error of the linear terms. For all of our simulations, the amplitudes
in the packet are of the order of $10^{-2}$ or larger.  Therefore the roundoff
is affecting only the amplitudes far in the exponential tails. We changed the
calculation to single precision, for which $a=8$, but we did not observe any
qualitative difference in our results. For single precision the nonlinear
terms will be affected by roundoff errors when $\sup_l|\psi_l|^2 < 10^{-4}$,
which is still realized only in the exponential tails.  We note, that the
times at which the roundoff errors affect the packet modes correspond to
$t\sim 10^{80}$ for $a=16$ and $t \sim 10^{30}$ for $a=8$ which are obviously
not accessible with our computation schemes.

Therefore we implemented a brute force roundoff scheme: after each time step of
 integration we take the distributions and perform a roundoff at a
prescribed digit $a=1,2,3,4,\ldots$.  We expect therefore to reduce the
time at which roundoff errors will become visible, in order to observe that
effect within the time window accessible by our computations.  Indeed, we find
that strong fluctuations in the conserved quantities set in at a time $t_{r}$
which decreases with decreasing $a$. In particular for the DNLS we find $t_r
\approx 10^3,10^5,10^7$ for $a=1,2,3$, and for the KG model we find $t_r
\approx 10^3,10^5,10^8$ for $a=1,2,3$.  When monitoring the second moment and
the participation number, we also find strong deviations from the above
results at times $t>t_r$. For $a \geq 4$ we do not observe
any significant change in the data. Therefore we conclude, that the roundoff
errors with double (or even single) precision are not affecting our results.

\section{Spreading mechanisms}
\label{sec.spreading}

We can think of two possible mechanisms of wave packet spreading. A NM with
index $\mu$ in a layer of width $\overline{p}$ in the cold exterior, which
borders the packet, is either incoherently {\sl heated} by the packet, or {\sl
resonantly excited} by some particular NM from a layer with width
$\overline{p}$ inside the packet. Heating here implies a (sub)diffusive
spreading of energy. Note that the numerical results yield subdiffusion,
supporting the nonballistic diffusive heating mechanism.

For heating to work, the packet modes $\phi_{\nu}(t)$ should contain a part
$\phi_{\nu}^{c}(t)$, having a continuous frequency spectrum (similar to a
white noise), in addition to a regular part $\phi_{\nu}^{r}(t)$ of pure point
frequency spectrum:
\begin{equation}
\phi_{\nu}(t) = \phi_{\nu}^{r}(t) + \phi_{\nu}^{c}(t)\;.
\label{cpp}
\end{equation}
Therefore at least some NMs of the packet should evolve chaotically in time.
The more the packet spreads, the less the mode amplitudes in the packet
become. Therefore its dynamics should become more and more regular, implying $
\lim_{t\rightarrow \infty} \phi_{\nu}^{c}(t)/\phi_{\nu}^{r}(t) \rightarrow 0$.

\subsection{Are all packet modes chaotic?}
\label{sec:dephase}

In Ref. \cite{PS08} it was assumed that all NMs in the packet are chaotic, and
their phases can be assumed to be random at all times.  At variance to the
above expectation, it follows that $\phi_{\nu}^{r}(t)=0$, or at least the
ratio $\phi_{\nu}^{c}(t)/\phi_{\nu}^{r}(t)$ is constant on average.
Consequently $|\phi_{\nu}^{c}(t)| \sim n^{1/2}$ where $n$ is the average norm
density in the packet.

According to (\ref{NMeq}) the heating of the exterior mode should evolve as $i
\dot{\phi}_{\mu} \approx \lambda_{\mu} \phi_{\mu} + \beta n^{3/2} f(t)$ where
$\langle f(t) f(t') \rangle = \delta(t-t')$ ensures that $f(t)$ has a
continuous frequency spectrum.  Then the exterior NM  increases its norm
according to $|\phi_{\mu}|^2 \sim \beta^2 n^3 t$. The momentary diffusion rate
of the packet is given by the inverse time $T$ it needs to heat the exterior
mode up to the packet level: $D = 1/T \sim \beta^2 n^2$. The diffusion
equation $m_2 \sim D t$ yields $m_2 \sim \beta t^{1/2}$.  We tested the above
conclusions by enforcing decoherence of NM phases.  Each 100 time units on
average 50\% of the NMs were randomly chosen, and their phases were shifted by
$\pi$ (DNLS). For the KG case we changed the signs of the corresponding NM
momenta.  We obtain $m_2\sim t^{1/2}$ (see Fig.\ref{fig_dephase}).  Therefore,
when the NMs dephase completely, the exponent $\tilde{\alpha}=1/2$, {\sl
contradicting} numerical observations {\sl without dephasing}.  Thus, not all
NMs in the packet are chaotic, and dephasing is at best a partial outcome.

\subsection{Mode-mode resonances inside the packet}
\label{sec:m-m}

Chaos is a combined result of resonances and nonintegrability. Let us estimate
the number of resonant modes in the packet for the DNLS model.  Excluding
secular interactions, the amplitude of a NM with
$|\phi_{\nu}|^2 = n_{\nu}$ is modified by a triplet of other modes
$\vec{\mu}\equiv (\mu_1,\mu_2,\mu_3)$ in first order in $\beta$ as (\ref{NMeq})
\begin{equation}
|\phi_{\nu}^{(1)}| = \beta \sqrt{n_{\mu_1}n_{\mu_2}n_{\mu_3}}
R_{\nu,\vec{\mu}}^{-1}\;,\; R_{\nu,\vec{\mu}} \sim
\left|\frac{\vec{d\lambda}}{I_{\nu,\mu_1,\mu_2,\mu_3}}\right| \;,
\label{PERT1}
\end{equation}
where $\vec{d\lambda} =
\lambda_{\nu}+\lambda_{\mu_1}-\lambda_{\mu_2}-\lambda_{\mu_3}$.  The
perturbation approach breaks down, and resonances set in, when $\sqrt{n_{\nu}}
< |\phi_{\nu}^{(1)}|$.  Since all considered NMs belong to the packet, we
assume their norms to be equal to $n$ for what follows.  If three of the four
mode indices are identical, one is left with interacting NM pairs.  A
statistical analysis of the probability of resonant interaction was performed
in Ref. \cite{fks08}.  For small values of $n$ (i.e. when the packet has
spread over many NMs) the main contribution to resonances are due to rare
multipeak modes \cite{fks08}, with peak distances being larger than the
localization volume.  If two or none of the four mode indices are identical,
one is left with triplets and quadruplets of interacting NMs respectively. In
both cases the resonance conditions can be met at arbitrarily small values of
$n$ for NMs from one localization volume.

\begin{figure}
\includegraphics[angle=0,width=0.99\columnwidth]{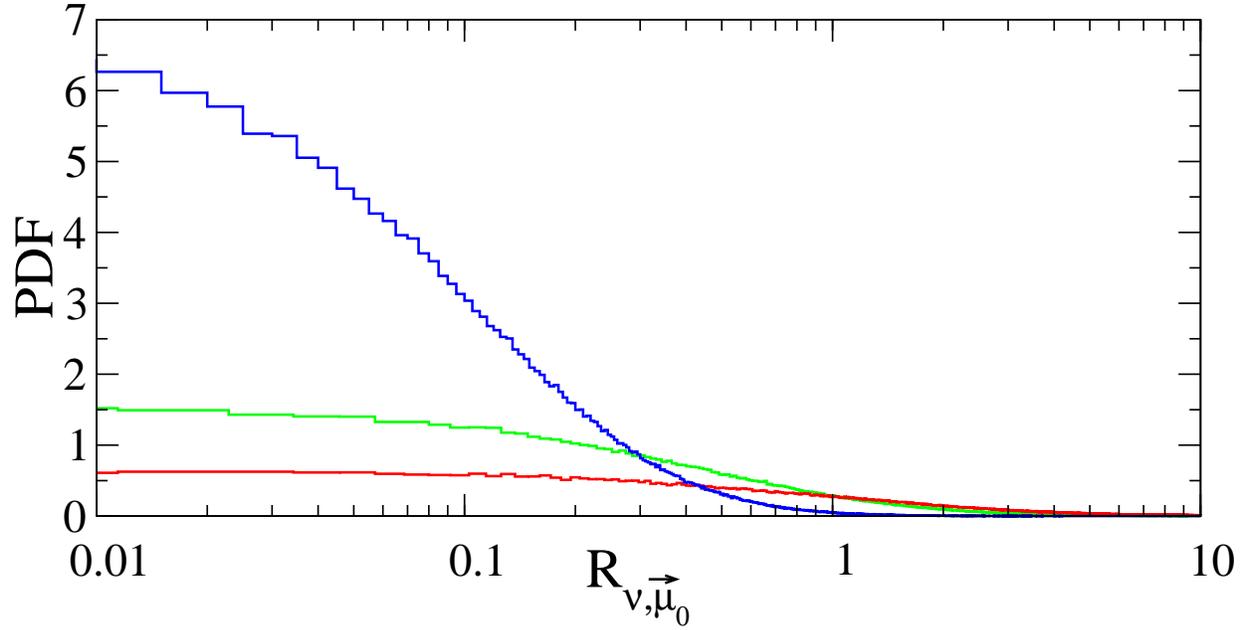}
\caption{(color online) Statistical properties of NMs of the DNLS model.
Probability densities $\mathcal{W}(R_{\nu,\vec{\mu}_0})$ of NMs being resonant
(see section \ref{sec:m-m} for details). Disorder strength $W=4,7,10$ (from
top to bottom).   }
\label{fig7}
\end{figure}

We perform a statistical numerical analysis for the quadruplet case. For a
given NM $\nu$ we obtain $ R_{\nu,\vec{\mu}_0} = \min_{\vec{\mu} }
R_{\nu,\vec{\mu}}$.  Collecting $R_{\nu,\vec{\mu}_0}$ for many $\nu$ and many
disorder realizations, we find the probability density distribution
$\mathcal{W}(R_{\nu,\vec{\mu}_0})$ (Fig.~\ref{fig7}).
The main result is that $\mathcal{W}(R_{\nu,\vec{\mu}_0} \rightarrow 0)
\rightarrow C(W) \neq 0$.  For the cases studied, the constant $C$ drops with
increasing disorder strength $W$. Similar results are found if pairs of
resonant NMs \cite{fks08} are analyzed, with the only difference that the
constant $C$ is reduced e.g. by a factor of 30 for $W=4$.

The probability $\mathcal{P}$ for a mode, which is excited to a norm $n$ (the
average norm density in the packet), to be resonant with at least one triplet
of other modes at a given value of the interaction parameter $\beta$ is given
by
\begin{equation}
\mathcal{P} = \int_0^{\beta n} \mathcal{W}(x) {\rm d}x\;,
\label{resprob}
\end{equation}
with $x$ denoting $R_{\nu,\vec{\mu}_0}$. For $\beta n \ll 1$ it follows
\begin{equation}
\mathcal{P} \approx C \beta n\;.
\label{resprobas}
\end{equation}
Therefore the probability for a mode in the packet to be resonant is
proportional to $C \beta n$.  On average the number of resonant modes in the
packet is constant, proportional to $C \beta$, and their fraction within the
packet is $\sim C \beta n$.  Since packet mode amplitudes fluctuate in
general, averaging is meant both over the packet, and over suitably long time
windows (yet short compared to the momentary inverse packet growth rate).  We
conclude, that the continuous frequency part of the dynamics of a packet mode
is scaled down by $C \beta n$, compared to the case when all NMs would be
chaotic. It follows that $\phi_{\nu}^{c}(t) / \phi_{\nu}^{r}(t) \sim C \beta
n$. As expected initially, the chaotic part in the dynamics of packet modes
becomes the weaker the more the packet spreads, and the packet dynamics
becomes more and more regular in the limit of large times.  Therefore the
chaotic component $\phi_{\nu}^{c}(t) \ll \phi_{\nu}^{r}(t)$ is a small
parameter.

Expanding the term $|\phi_{\nu}|^2\phi_{\nu}$ to first order in
$\phi_{\nu}^{c}(t)$, the heating of the exterior mode should evolve according
to $i \dot{\phi}_{\mu} \approx \lambda_{\mu} \phi_{\mu} + C \beta^2 n^{5/2}
f(t)$.  It follows $|\phi_{\mu}|^2 \sim C^2 \beta^4 n^5 t$, and the rate $D =
1/T \sim C^2 \beta^4 n^4$ (cf.~the prediction (\ref{diffcoef})). The
diffusion equation $m_2 \sim D t$ yields
\begin{equation}
m_2 \sim C^{2/3} \beta^{4/3} t^{\alpha}\;,\; \alpha = 1/3\;.
\label{subdif}
\end{equation}
The predicted exponent $\alpha=1/3$ is close to the numerically observed one,
as we discussed in section \ref{sec:spread}.

\subsection{Resonant spreading?}

Finally we consider the process of resonant excitation of an exterior mode by
a mode from the packet. The number of packet modes in a layer of the width of
the localization volume at the edge, which are resonant with a cold exterior
mode, will be proportional to $\beta n$. After long enough spreading $\beta n
\ll 1$.  On average there will be no mode inside the packet, which could
efficiently resonate with an exterior mode. Therefore, resonant growth can be
excluded.

\section{Summary and discussion}

We studied the spreading of wave packets in disordered one--dimensional
nonlinear chains. In particular we considered two systems, namely the DNLS
model (\ref{RDNLS}) and the quartic KG system (\ref{RQKG}). The linear parts
of these two models are equivalent in the sense that they correspond to the
same eigenvalue problem (\ref{EVequation}).

We predicted theoretically and verified numerically the existence of three
different dynamical behaviors depending on the relation of the nonlinear
frequency shift $\delta$ (which is proportional to the system's nonlinearity)
with the average spacing $\overline{\Delta \lambda}$ of eigenfrequencies and
the spectrum width $\Delta$ ($\overline{\Delta \lambda} \leq \Delta)$ of the
linear system. The dynamics for small nonlinearities ($\delta <
\overline{\Delta \lambda}$) is characterized by localization as a transient,
with subsequent subdiffusion (regime I). For intermediate values of the
nonlinearity $ \overline{\Delta \lambda} < \delta <\Delta$, and the wave
packets exhibit immediate subdiffusion (regime II). In this case, the second
moment $m_2$ and the participation number $P$ increase in time following the
power laws $m_2 \sim t^{\alpha}$, $P \sim t^{\alpha/2}$. Assuming that the
spreading is due to an incoherent excitation of the cold exterior, induced by
the chaotic behavior of the wave packet, we predicted $\alpha=1/3$. Finally,
for even higher nonlinearities ($\delta > \Delta$) a large part of the wave
packet is selftrapped, while the rest subdiffuses (regime III). In this case
$P$ remains practically constant, while $m_2 \sim t^{\alpha}$. The overall
picture is schematically presented in Fig.~\ref{fig_param} both for the DNLS
and the KG model.

The compactness index $\zeta=P^2/m_2$, which measures the sparseness of wave
packets, exhibits different behaviors for the three dynamical regimes. In
particular, the behavior of $\zeta$ for wave packets in regime II imply that
these wave packets spread but do not become more sparse.

For large values of the disorder strength $W$ and/or strong nonlinearity the
intermediate regime II effectively disappears, and the evolution will start
either in regime I, or in regime III. In regime I the detrapping times
increase with further increase of $W$. In regime III the fraction of the
wave packet which spreads decreases with increasing nonlinearity. Therefore,
large values of $W$ and/or nonlinearity will not allow for an observation of
the destruction of Anderson localization on time scales which are bounded from
above by practical computational limitations.

The subdiffusive spreading is universal, i.~e.~the exponent $\alpha$ is
independent of the nonlinearity's strength ($\beta$ for the DNLS model and
energy $E$ for the KG one) and $W$, which are only affecting the prefactor in
(\ref{subdif}).  Excluding selftrapping, any nonzero nonlinearity strength
$\beta$ will completely delocalize the wave packet and destroy Anderson
localization.  The exponent $\alpha$ is determined solely by the degree of
nonlinearity, which defines the type of overlap integral to be considered in
(\ref{PERT1}), and by the stiffness of the spectrum $\{ \lambda_{\nu} \}$. Our
numerical computations confirmed the prediction $\alpha=1/3$ in the case of
single site and of nonlocal homogeneous excitation. In the case of single mode
excitations the three different regimes were also detected. The numerically
computed exponents $\alpha$ get slightly larger values than $1/3$, exhibiting
also a small dependence on the strength of nonlinearity. This discrepancy
between the two cases in not clearly understood.

We studied the statistics of detrapping times $\tau_d$ for regime I. We
provided numerical evidences for the validity of the conjectured dependence of
$\tau_d$ on the nonlinearity strength and on the average norm density of the
excited NMs given in Eq.~(\ref{diffcoef}). It is worth mentioning that,
distributing the energy of a single site excitation belonging to regime II
over more sites results in a time dependence of $m_2$ and $P$ similar to
regime I. In addition, considering as initial condition the profile of a
single site excitation in regime II at some latter time $t_d$, we observe a
dynamical evolution of the type of regime I where the detrapping time is
$\tau_d \approx t_d$.

The spreading of the wave packet is due to weak but nonzero chaotic dynamics
inside the packet.  It is natural to expect such a dynamics, since the
considered systems are nonintegrable.  If instead an integrable system is
considered, Anderson localization will not be destroyed.  Indeed, consider a
Hamiltonian in NM representation using actions $J_{\nu}$ and angles
$\theta_{\nu}$ as coordinates:
\begin{equation}
\mathcal{H}_{int} = \sum_{\nu} \lambda_{\nu} J_{\nu} + \beta
\sum_{\nu_1,\nu_2,\nu_3,\nu_4} I_{\nu_1,\nu_2,\nu_3,\nu_4} \sqrt{ J_{\nu_1}
J_{\nu_1} J_{\nu_1} J_{\nu_1}}\;.
\label{int1}
\end{equation}
We assume that the set of eigenfrequencies $\{ \lambda_{\nu} \}$ and the
overlap integrals $I_{\nu_1,\nu_2,\nu_3,\nu_4}$ are identical with those
describing the DNLS model (\ref{NMeq}), (\ref{OVERLAP}). The equations of
motion $\dot{J}_{\nu} = -\partial \mathcal{H}_{int}/ \partial \theta_{\nu}$
and $\dot{\theta}_{\nu} = \partial \mathcal{H}_{int}/ \partial J_{\nu}$ yield
$\dot{J}_{\nu} = 0$ since the integrable Hamiltonian (\ref{int1}) depends only
on the actions.  Therefore, any localized initial condition
(e.~g.~$J_{\nu}(t=0) \propto \delta_{\nu,\nu_0}$) will stay localized, since
actions of modes which are at large distances will never get  excited.
Thus, the observed spreading of wave packets, which we studied in detail in
the present work, is entirely due to the nonintegrability of the considered
models, at variance to (\ref{int1}).

The more the wave packet spreads, the weaker the resonances
become. Corresponding structures (chaotic layers) in phase space become
thinner and thinner.  Consider quantum many-body systems.  Classical phase
space structures which are finer than the action quantization induced grid
become irrelevant. Therefore we may speculate, that the wave packet will stop
spreading for a quantum many-body system at some point for zero temperature,
but also for temperatures below some finite threshold. These expectations are
very close to rigorous results for interacting fermions in disordered systems
\cite{dmbilabla06}.

In our study we considered initial conditions exciting NMs whose eigenvalues
are located close to the center of the frequency band. Thus, the evolution of
the system does not significantly depend on the sign of nonlinearity. In
contrast, when one excites eigenstates with frequencies near the band edges, a
rather weak nonlinearity might lead either to selftrapping or to the weak
nonlinear regime depending on the sign of nonlinearity. Such examples were
presented in \cite{Exp2} where NMs close to the edges of the band exhibit
different dynamical behaviors, i.~e.~one becomes more localized as the
nonlinearity was switched on, while the other tends to delocalize.  If a
spatially continuous system is considered, then a proper choice of the sign of
nonlinearity prohibits selftrapping (so-called defocusing nonlinearity,
corresponding to repulsive two-body interactions).  For such a case, regime
III ceases to exist, and localization is expected to be destroyed
irrespectively of the strength of nonlinearity.

\begin{acknowledgments}
 We thank B.~L.~Altshuler, S.~Aubry, G.~Kopidakis and R.~Schilling
 for useful discussions.
\end{acknowledgments}

\appendix
\section{The SABA$_2$ and SBAB$_2$ symplectic integrators }
\label{sec:saba2c}

In \cite{LR01} a family of symplectic integrators which involve only forward
integration steps was proposed. These integrators were adapted for
integrations of perturbed Hamiltonians of the form
\begin{equation}
H=A+\epsilon B,
\label{eq:hamab}
\end{equation}
where both $A$ and $B$ are integrable and $\epsilon$ is a parameter. We
briefly recall here their main properties focusing our attention on two
particular members of the family of integrators presented in \cite{LR01},
namely the SABA$_2$ and SBAB$_2$ integrators. These integrators have already
proved to be very efficient for the numerical study of astronomical
\cite{LR01}, as well as accelerator models \cite{accel}.

Consider a Hamiltonian system of $N$ degrees of freedom having a Hamiltonian
$H(\vec{p}, \vec{u})$, with $\vec{p}=(p_1, \ldots, p_N)$, $\vec{u}=(u_1,
\ldots, u_N)$ where $u_l$ and $p_l$, $l=1,\ldots,N$, are the generalized
coordinates and momenta respectively. An orbit of this system is defined by a
vector $\vec{x}(t)=(x_1(t), \ldots, x_{2N}(t))$, with $x_l=p_l$,
$x_{l+N}=u_l$, $l=1,\ldots,N$. This orbit is a solution of Hamilton's
equations of motion:
\begin{equation}
\frac{d \vec{p_l}} {dt}= - \frac{\partial H}{\partial \vec{u_l}}\,\,\, ,
\,\,\, \frac{d \vec{u_l}}{dt}= \frac{\partial H}{\partial \vec{p_l}}\,\,\,
,\,\,\, l=1,\ldots,N,
\label{eq:Hameq}
\end{equation}
where $t$ is the independent variable, namely the time. Defining the Poisson
bracket of functions $f(\vec{p},\vec{u})$, $g(\vec{p},\vec{u})$ by:
\begin{equation}
\{ f,g\}=\sum_{l=1}^{N} \left( \frac{\partial f}{\partial p_l} \frac{\partial
g}{\partial u_l} - \frac{\partial f}{\partial u_l} \frac{\partial g}{\partial
p_l}\right) ,
\label{eq:Poisson}
\end{equation}
the Hamilton's equations of motion take the form:
\begin{equation}
\frac{d \vec{x}} {dt}= \{H,\vec{x}\} = L_H \vec{x},
\label{eq:Hameq_x}
\end{equation}
where $L_H$ is the differential operator defined by $L_{\chi}f=\{\chi, f \}$.
The solution of Eq.~(\ref{eq:Hameq_x}), for initial conditions
$\vec{x}(0)=\vec{x}_0$, is formally written as:
\begin{equation}
\vec{x}(t)=\sum_{n\geq 0} \frac{t^n}{n!} L_H^n \vec{x}_0=e^{t L_H}
\vec{x}_0.
\label{eq:Ham_sol}
\end{equation}

A symplectic scheme for integrating (\ref{eq:Hameq_x}) from time $t$ to time
$t+\tau$ consists of approximating, in a symplectic way, the operator $
e^{\tau L_H}= e^{\tau(L_A+L_{\epsilon B} ) } $ by an integrator of $j$ steps
involving products of $e^{c_i \tau L_A}$ and $e^{d_i \tau L_{\epsilon B}}$,
$i=1,2,\ldots, j$, which are exact integrations over times $c_i \tau$ and $d_i
\tau$ of the integrable Hamiltonians $A$ and $B$. The constants $c_i$, $d_i$,
are chosen so that to increase the order of the remainder of this
approximation.

For the SABA$_2$ integrator we get:
\begin{equation}
\mbox{SABA}_2= e^{c_1 \tau L_A} e^{d_1 \tau L_{\epsilon B}} e^{c_2 \tau L_A}
e^{d_1 \tau L_{\epsilon B}} e^{c_1 \tau L_A},
\label{eq:saba2}
\end{equation}
with $c_1=\frac{1}{2}\left( 1- \frac{1}{\sqrt{3}} \right)$,
$c_2=\frac{1}{\sqrt{3}}$, $d_1= \frac{1}{2}$, while the SBAB$_2$ integrator is
given by
\begin{equation}
\mbox{SBAB}_2= e^{d_1 \tau L_{\epsilon B}} e^{c_2 \tau L_A}
e^{d_2 \tau L_{\epsilon B}} e^{c_2 \tau L_A} e^{d_1
\tau L_{\epsilon B}},
\label{eq:sbab2}
\end{equation}
with $c_2=\frac{1}{2}$, $d_1= \frac{1}{6}$, $d_2= \frac{2}{3}$.  Using
these integrators we are actually approximating the dynamical behavior of the
real Hamiltonian $A+\epsilon B$ by a Hamiltonian $\widetilde{H} = A+\epsilon B
+ \mathrm{\cal{O}}(\tau^4 \epsilon +\tau^2 \epsilon ^2)$, i.~e.~we introduce
an error term of the order $\tau^4 \epsilon +\tau^2 \epsilon ^2$.

The accuracy of the  SABA$_2$ (or SBAB$_2$) integrator can be improved when the
term $C=\{\{A,B\},B\}$ leads to an integrable system, as in the common
situation of $A$ being quadratic in momenta $\vec{p}$ and $B$ depending only
on positions $\vec{u}$. In this case, two corrector terms of small backward
steps can be added to the integrator SABA$_2$
\begin{equation}
\mbox{SABA}_2\mbox{C}= e^{- \tau^3 \epsilon^2 \frac{g}{2}
L_{C}}(\mbox{SABA}_2)e^{- \tau^3 \epsilon^2 \frac{g}{2} L_{C}}.
\label{eq:saba2c}
\end{equation}
A similar expression is valid also for SBAB$_2$. The value of $g$ was chosen
in order to eliminate the $\tau^2 \epsilon^2$ dependence of the remainder
which becomes of order $\mathrm{\cal{O}}(\tau^4 \epsilon +\tau^4 \epsilon
^2)$. In particular we have $g=(2-\sqrt{3})/24$ for SABA$_2$ and
$g=\frac{1}{72}$ for SBAB$_2$.  We note that the SABA$_2$ and SBAB$_2$
integrators involve only forward steps which increases their numerical
stability, while, the addition of the corrector results to better accuracy of
the schemes, introducing simultaneously a small backward step.

\subsection{Integration of the KG lattice}
\label{sec:saba2c_KG}

Hamiltonian (\ref{RQKG}) is suitable for the implementation of the SABA$_2$C
integration scheme since it attains the form (\ref{eq:hamab}) with:
\begin{equation}
\begin{array}{lll}
A & \equiv & \displaystyle \sum_{l=1}^N \frac{p_{l}^2}{2},\\ B & \equiv &
 \displaystyle \sum_{l=0}^N \frac{\tilde{\epsilon}_{l}}{2} u_{l}^2 +
 \frac{1}{4} u_{l}^{4}+\frac{1}{2W}(u_{l+1}-u_l)^2, \\ \displaystyle \epsilon
 &= &1,
\end{array}
\label{eq:ham_equiv}
\end{equation}
where $N$ is the number of anharmonic oscillators. The operators $e^{\tau
L_A}$, $e^{\tau L_B}$, $e^{\tau L_C}$, which propagate the set of initial
conditions $(u_l, p_l)$ at time $t$, to their final values $(u'_l, p'_l)$ at
time $t+\tau$, $l=1,2,\ldots,N$ are:
\begin{equation}
e^{\tau L_A}: \left\{ \begin{array}{lll} u'_l & = & p_l \tau + u_l\\ p'_l & =
& p_l \\
\end{array}\right. ,
\label{eq:LA}
\end{equation}
\begin{widetext}
\begin{equation}
e^{\tau L_B}: \left\{ \begin{array}{lll} u'_l & = & u_l \\ p'_l & = & \left[
-u_l \left( \tilde{\epsilon}_l+u_l^2 \right) +\dfrac{1}{W} \left( u_{l-1}+
u_{l+1}-2 u_l\right) \right] \tau + p_l \\
\end{array}\right. ,
\label{eq:LB}
\end{equation}
\begin{equation}
e^{\tau L_C}: \left\{ \begin{array}{lll}
u'_l  & = &  u_l  \\
p'_1  & = & 2 \left\lbrace  \left( \dfrac{2}{W}+\tilde{\epsilon}_1 + 3 u_1^2 \right)
\left[  -u_1 \left( \tilde{\epsilon}_1+u_1^2 \right) +\dfrac{1}{W} \left( u_{2}-2 u_1\right) \right] \right. \\
& & \left. + \dfrac{1}{W} \left[  u_{2} \left( \tilde{\epsilon}_{2}+u_{2}^2 \right) -\dfrac{1}{W} \left( u_{3}+ u_{1}-2 u_{2}\right) \right]\right\rbrace \tau + p_1  \\
p'_l  & = & 2 \left\lbrace \dfrac{1}{W} \left[  u_{l-1} \left( \tilde{\epsilon}_{l-1}+u_{l-1}^2 \right) -\dfrac{1}{W} \left( u_{l-2}+ u_{l}-2 u_{l-1}\right) \right] \right.    \\
& &  + \left[ \dfrac{2}{W}+\tilde{\epsilon}_l + 3 u_l^2 \right]
\left[  -u_l \left( \tilde{\epsilon}_l+u_l^2 \right) +\dfrac{1}{W} \left( u_{l-1}+ u_{l+1}-2 u_l\right) \right]  \\
& & \left. + \dfrac{1}{W} \left[  u_{l+1} \left( \tilde{\epsilon}_{l+1}+u_{l+1}^2 \right) -\dfrac{1}{W} \left( u_{l+2}+ u_{l}-2 u_{l+1}\right) \right]\right\rbrace \tau + p_l, \,\,\,\,\, \mbox{for}\,\,\, l=2,3,\ldots,N-1 \\
p'_N  & = & 2 \left\lbrace \dfrac{1}{W} \left[  u_{N-1} \left( \tilde{\epsilon}_{N-1}+u_{N-1}^2 \right) -\dfrac{1}{W} \left( u_{N-2}+ u_{N}-2 u_{N-1}\right) \right] \right.    \\
& & \left. + \left( \dfrac{2}{W}+\tilde{\epsilon}_N + 3 u_N^2 \right)
\left[  -u_N \left( \tilde{\epsilon}_N+u_N^2 \right) +\dfrac{1}{W} \left( u_{N-1}-2 u_N\right) \right] \right\rbrace \tau + p_N  \\
\end{array}\right. ,
\label{eq:LC}
\end{equation}
since
\begin{equation}
C=\sum_{l=1}^N \left[u_l \left( \tilde{\epsilon}_1 +u_l^2\right) -\dfrac{1}{W}  \left( u_{l-1}+ u_{l+1}-2 u_{l}\right) \right]^2 ,
\label{eq:ABB}
\end{equation}
and $u_0=u_{N+1}\equiv0$.
\end{widetext}

\subsection{Integration of the DNLS lattice}
\label{sec:sbab_DNLS}

We use the SBAB$_2$ integrator scheme to integrate the equations of motion
(\ref{RDNLS-EOM}), by splitting the DNLS Hamiltonian (\ref{RDNLS}) as
\begin{equation}
\begin{array}{lll}
A & \equiv & \displaystyle -\sum_{l=1}^{N}
 (\psi_{l+1}\psi_l^*+\psi_{l+1}^*\psi_l),\\ B & \equiv & \displaystyle
 \sum_{l=1}^{N}\epsilon_{l} |\psi_{l}|^2+\frac{\beta}{2} |\psi_{l}|^{4}, \\
 \displaystyle \epsilon &= &1,
\end{array}
\label{eq:ham_equiv_DNLS}
\end{equation}
with $N$ being the number of lattice sites.
The action of the operator
$e^{\tau L_A}$ on $\psi_l$, $l=1,2,\ldots,N$ at time $t$ leads to the
computation of $\psi'_l$ at time $t+\tau$, and includes three steps: a) the
transformation of the wavefunction from the real ($\psi_l$) to the Fourier
($\varphi_q$) space, through a Fast Fourier transform (FFT), b) a rotation
of $\varphi_q$, and c) the inverse FFT of the wavefunction $\varphi'_q$
evaluated at the previous step, i.~e.
\begin{equation}
e^{\tau L_A}: \left\{ \begin{array}{lll}
{\varphi_q}&=&\sum_{m=1}^{N}\psi_{m} e^{2\pi i q(m-1)/N} \\\\
{\varphi'_q}&=&{\varphi_q} e^{2i \cos\left(2\pi (q-1)/N\right)\tau} \\\\
\psi'_l&=&\dfrac{1}{N}\sum_{q=1}^{N}\varphi'_q e^{-2\pi i l(q-1)/N}
\end{array}\right. .
\label{eq:LA_DNLS}
\end{equation}
On the other hand, the action of $e^{\tau L_B}$ on $\psi_l$ reduces to a
simple rotation in real space, namely
\begin{equation}
e^{\tau L_B}: \left\{ \begin{array}{lll}
\psi'_l=\psi_l e^{-i(\epsilon_l+\beta|\psi_l|^2)\tau}\end{array}\right..
\label{eq:LB_DNLS}
\end{equation}
Note that for the DNLS model we do not apply the two corrector steps since the
term $C=\{\{A,B\},B\}$ does not lead to an easily solvable system.
\\
\\
\\
\\
\\


\end{document}